\documentclass[a4paper,11pt]{article}
\usepackage{amsmath, amsthm, amssymb}
\usepackage{multicol}
\usepackage{graphicx}
\usepackage{epsfig}
\usepackage{subfigure}
\usepackage{psfrag}
\usepackage{footnote,ctable}
\newcommand{\app}[1]{
  \refstepcounter{section}
  \setcounter{equation}{0}
  \mbox{\ }\smallskip
  {\begin{flushleft}
  \bf \large Appendix \thesection.\ {#1}
  \end{flushleft}}
  }

\newcommand{\be}{\begin{equation}}
\newcommand{\ee}{\end{equation}}

\begin{document}

\title{
 \huge Single and two-scale sharp-interface models for concrete carbonation -- Asymptotics and numerical approximation \vspace*{0.6cm}}
\author
{ { Jonathan D. Evans}\footnote{Department of Mathematical Sciences,
 University of Bath, Bath, BA2 7AY, UK.  E-mail:
masjde@maths.bath.ac.uk}  { Andrea Fern\'{a}ndez}\footnote{Department of Mathematical Sciences,
 University of Bath, Bath, BA2 7AY, UK.  E-mail:
af243@bath.ac.uk}  { Adrian Muntean}\footnote{Center for Analysis, Scientific Computing and Applications (CASA), Department of Mathematics and Computer
Science, Institute for Complex Molecular Systems (ICMS), Eindhoven University of Technology, PO Box 513, 5600 MB Eindhoven, The Netherlands. E-mail: a.muntean@tue.nl}
}

\date{\today}

\maketitle \vspace*{-0.5cm}

\thispagestyle{empty}

\begin{abstract}
We investigate the fast-reaction asymptotics for a  one-dimensional reaction-diffusion (RD) system  describing the penetration of the
carbonation reaction in concrete. The technique of matched-asymptotics is used to show that the RD system leads to two distinct classes of  sharp-interface models,
that correspond to different scalings in a small parameter $\epsilon$ representing the fast-reaction. Here $\epsilon$ is the ratio between the characteristic scale of the diffusion of the fastest species and the one of the carbonation reaction.  We explore three conceptually different scaling regimes (in terms of  $\epsilon$) of the effective diffusivities of the driving chemical species.  The limiting models include one-phase and two-phase generalised Stefan moving-boundary problems as well as a nonstandard two-scale (micro-macro) moving-boundary problem -- the main result of the paper. Numerical results, supporting the asymptotics, illustrate the behavior of the concentration profiles for relevant parameter regimes.
\end{abstract}

\vskip0.8mm
\textit{Abbreviations: FBP -- free boundary problem; PDE -- partial differential equation}

\setcounter{equation}{0}
\section{Introduction}
\label{intro}

Carbonation reactions alter the concrete's pore geometry in a difficult-to-control fashion.  A good theoretical multiscale understanding of the evolution of carbonation reactions in such a complex multiphase material is vital towards obtaining accurate predictions of the durability of large concrete structures  (see, for example,  \cite{Kropp,Kim}). Having a good estimate on the life service of motorways, bridges, sewage pipe systems (etc.) can save significant amounts of money and energy yearly -- hence the growing multidisciplinary research interest in this topic.

\subsection{Background}

\begin{figure}[h!]
\centering
\includegraphics[scale=0.75]{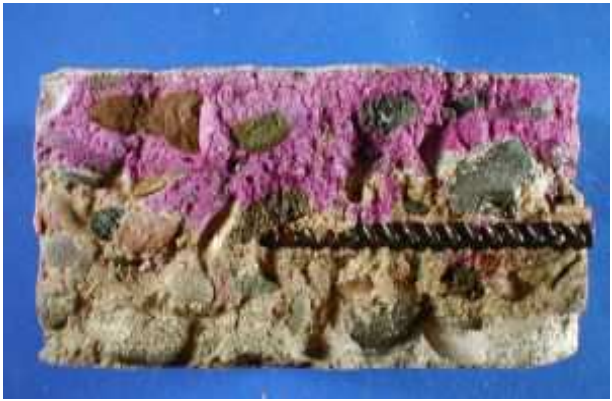}
\caption{Cross-section of a reinforced concrete block. Pink indicates low pH while colourless indicates a high one.}
\label{Carbpic}
\end{figure}

From the point of view of the involved geochemistry, the process of concrete carbonation mainly involves the reaction of atmospheric carbon dioxide with calcium hydroxide found in the pore solution to form water and calcium carbonate. The pore solution is highly alkaline (pH $\sim$ 14) and as soon as the pH level decreases the microscopic oxide layer at the steel reinforcements disappears and the steel can corrode. This corrosion causes the durability of the structure to reduce dramatically. Figure \ref{Carbpic} shows how the pH levels of the concrete block drop dramatically. For more details on concrete carbonation and its relevance with respect to corrosion and durability issues, we refer the reader to ~\cite{PVF}, ~\cite{Kropp} and ~\cite{BBC} as well as to the references cited therein.

Here, we focus on the dominant carbonation reaction mechanism, namely
\begin{equation}
\mathrm{CO}_{2}{(g\rightarrow aq)} + \mathrm{Ca(OH)}_{2}{(s\rightarrow aq)}  \rightarrow \mathrm{CaCO}_{3}{(aq\rightarrow s)}+ \mathrm{H}_2\mathrm{O} .\label{carbreact}
\end{equation}
In other words, atmospheric carbon dioxide diffuses through the unsaturated concrete matrix and dissolves into the pore water while the calcium hydroxide is available in the pore solution by dissolution from the solid matrix. Free water and calcium carbonate are the main reaction products in  \eqref{carbreact}. ${\rm CaCO_3}$ quickly precipitates into the solid matrix, while the water diffuses through the material very much influenced by eventual wetting and deleting cycles. As a consequence of so many reaction and transport mechanisms, the available macroscopic models capturing carbonation in concrete are complex and involve a large number of parameters. For this reason, although the mechanisms are rather well understood, the predictability power of such models is questionable. The challenge is to find minimal models that can accurately predict ${\rm CO_2}$ penetration depths in concrete.

Our objective is to understand how fast the reaction \eqref{carbreact} moves into the material, driven by ${\rm CO_2}$ diffusion. In other words, we want to use a detailed reaction layer analysis to determine the speed of growth of the macroscopic carbonated phase (the pink region in Figure \ref{Carbpic}). The output of our approach/models consists of carbonation penetration depths and profiles of active concentrations.
\newline
In this context,  the open question is:

\begin{center} (Q)
\quad {\em Which boundary conditions need to be imposed at the reaction interface to describe its correct motion?}\end{center}
(Q) has been recently addressed, for instance,  in \cite{Muntean_thesis, MB09, Ai09, MB09_fast, Mun09_RWA, Ai10, Mun_mec}, and \cite{MBK11}, where the authors indicate that  the answer to (Q) seem to be rather well understood in one-space dimension. However, the $2D$ and $3D$ cases are comparatively untouched. The reason is simple: For the moment, one does not know  how the reaction front [hosting (\ref{carbreact})] feels the presence of corners or more complex geometrical shapes. Furthermore, the mathematical theory of FBPs in more space dimensions [and posed in heterogeneous media] is less understood. Rigorous results are mostly available for standard FBPs, relying heavily on a lot of  {\em a priori} prescribed regularity for the  moving boundary (sharp interface). For practical purposes, we need robust models [$\epsilon$-approximations of sharp-interface models for carbonation] that are appropriate in any space dimension. Such models should not require any special regularity requirements on the sharp interface and should be able to capture as $\epsilon\to 0$ the correct moving interface conditions. This is the place where our paper contributes. The research reported here is preliminary, with our main asymptotic results emphasizing a new, non-standard two-scale sharp-interface model. Many associated fundamental and practical issues are currently open (see section \ref{Discussion}) and deserve further study.

\subsection{Approach}

We introduce a basic macroscopic reaction-diffusion model to describe the aqueous chemistry and transport involved in \eqref{carbreact}. To be more precise, we only consider the concentration of the reactants ${\rm CO_2(aq)}$ and ${\rm Ca(OH)_2(aq)}$ which we allow both to diffuse. Furthermore, we assume that the diffusion coefficients can be concentration dependent and also may vary with the reaction rate. We then use the technique of matched-asymptotics~\cite{BO} to perform a detailed analysis of the reaction layer for various physically relevant scaling regimes. Additionally, we use COMSOL to produce numerical results of the original system that corroborate the asymptotic results obtained for the various scaling cases.

Since we look to carbonation in high Thiele moduli regimes, the rate at which the concentrations diffuse is typically much slower than that of the reaction. Thus we focus on the {\em fast reaction limit} $\epsilon\to 0$, where $\epsilon>0$ is a scale parameter that will be explained in the next section. Assuming strong time scales separation, the reaction is expected to be in equilibrium except for  an interior narrow layer\footnote{Interestingly, the strong separation of the characteristic time scales of diffusion and reaction lead to the presence of two distinct (macroscopic) characteristic space scales that  relate to the typical sizes of the layers hosting diffusion and reaction.} where virtually all the reaction takes place. The structure of this localized reaction zone is sensitive to the behaviour of the diffusion coefficients. For this reason, we investigate three different possible situations in regard to the diffusivities.  These lead to the derivation of conceptually different sharp-interface models.  We see three main types; namely, one and two phase Stefan like problems with zero latent heat and two scale micro-macro free boundary problems. The latter is a completely new type of model in which the speed of the reaction interface is updated from a smaller length scale.

The broad spectrum of different models showed here depicts not only the behaviour of the concentrations for many possible scenarios of carbonation, but they can also be related to other types of aggressive reactions like redox scenarios \cite{Hagan}, silicon-oxidation ~\cite{EK}, sulfate attack \cite{natalini,tasnim1,Niko} and combustion \cite{Victor}. The techniques used here are therefore applicable to a larger set of reaction-diffusion problems.

\subsection{Organization of the paper}

The macroscopic reaction-diffusion system  modelling carbonation is introduced in Section \ref{Model}. Performing our fast-reaction asymptotics, a first sharp-interface model is derived in Section \ref{FBP_first}. The bulk of the paper (Sections \ref{D0}, \ref{D1} and \ref{D2}) is devoted to the reaction layer analysis for three different diffusion coefficient regimes. The first regime (Section \ref{D0}) is that in which the diffusion coefficients remain order 1, which we refer to as slowly varying. This is the usual case which has received attention in the literature. The other regimes, allow the diffusion coefficients to change their orders of magnitude in terms of the parameter $\epsilon$, the diffusion coefficients then being termed rapidly varying. Thus, as a second regime (Section \ref{D1}) we consider only a rapidly varying CO$_2$ diffusivity, with the hydroxide diffusivity slowly varying. Finally, as a third regime (Section \ref{D2}), we consider the effects of rapidly varying CO$_2$ and hydroxide diffusivities simultaneously. The main results of the paper are the two-scale sharp interface models, which are summarised in \ref{SoltStefanProb}. The specific generalised Stefan and kinetic conditions on the moving boundary are summarised for the single and two rapidly varying cases in sections \ref{D1sum} and \ref{D2sum}, and should be compared to the conditions in the slowly varying case in section \ref{D0sum}. Each regime subdivides according to the size of a relative transport parameter, which measures the characteristic ratio of diffusivities and concentrations for CO$_2$ and hydroxide.

We close the paper with a discussion section around possible connections between the nonstandard two-scale models derived in this paper with the two-scale models obtained previously in the homogenization literature \cite{Hornung}.

\section{The model equations}\label{Model}

We consider the concrete occupying a one-dimensional slab geometry $\Omega:=]0,L[$. The $x$-axis be directed into the concrete with the surface $x=0$
being exposed to an external source of CO$_2$, whilst the surface $x=L$ is assumed impervious to all reaction species. We let $c=c(x,t)$ and
$h=h(x,t)$ denote the concentrations of CO$_{2_{(\mathrm{aq})}}$ and Ca(OH)$_2{_{(\mathrm{aq})}}$ respectively, expressed as moles per unit volume
(i.e. the intrinsic concentrations). We adopt the following set of reaction-diffusion equations,
\begin{align}
\hskip -1cm \text{In} \, 0<x<L,t>0  \hskip 1cm  & \frac{\partial c}{\partial t} = \frac{\partial}{\partial x}\left(D_c \frac{\partial c}{\partial x} \right) - R(c,h), \label{eq:dge1}\\
&\frac{\partial h}{\partial t} = \frac{\partial}{\partial x}\left(D_h \frac{\partial h}{\partial x}\right) - R(c,h) ;
\end{align}
with boundary conditions
\begin{align}
\text{on} \, x=0 \hskip 1cm & -D_c\frac{\partial c}{\partial x}=H^*(c^*-c), \; \frac{\partial h}{\partial x}=0 ; \label{eq:dbc1}\\
\text{on} \, x=L  \hskip 1cm & \frac{\partial c}{\partial x} = 0, \, \frac{\partial h}{\partial x} = 0 ;
\end{align}
and initial conditions
\be
       \text{at} \, t=0 \hskip 1cm c= c^* {c}_i(x), \, h= h^0 h_i(x).
\label{eq:dic}
\ee
The initial hydroxide concentration has a representative value (taken as maximum) denoted by the constant $h^0$, the most common situation being $h_i(x)=1,c_i(x)=0$ for $0\leq x \leq L$.
Here the concrete substrate has length L; R is
the carbonation rate with reaction rate coefficient $k$; $D_c$  and $D_h$ are the diffusion coefficients of CO$_2$ and Ca(OH)$_2$ respectively, which are not necessarily constant. Not only can the diffusion coefficients be concentration dependent\footnote{Consider the diffusion coefficient of CO$_2$, this may be expressed as $D_c=\hat{D}_cw\Phi(h)$, where $w$ is a tortuosity factor and
$\Phi(h)$ is the porosity which depends on the concentration of Ca(Oh)$_2$. This gives a justification for the CaOH$_2$ dependance of the
diffusivity. We refer the reader to \cite{Muntean_thesis} for more remarks in this direction.} but we also allow them to vary with the reaction rate. Motivation for this comes from the significant change that takes in the concrete matrix during carbonation. The main modelling assumption is thus to consider their dependence linked to $R(c,h)$, which for specificity may be taken in Arrhenius forms
\be
     D_i = D_i^0 \left( 1 - \exp \left( - \frac{\nu_i}{R_i} \right)  \right), \hskip 0.5cm i=c,h,
\label{eq:Di}
\ee
$\nu_c,\nu_h$ suitable constants and $R_c,R_h$ are based on partial reaction rates.
We note that neither the effective diffusivities nor the reaction rate coefficient depend on humidity. Also the transfer of
CO$_2$ from the air to water phase (and vice versa) and the dissolution of Ca(OH)$_2$ from the solid matrix to water phase (and viceversa) are in
local equilibrium i.e. all production terms by Henry's law vanish; see \cite{Danckwerts} for more details on molecular transfer across  water-air interfaces and on Henry's law. As such, from a modelling perspective, we may justify the Robin boundary
condition for the aqueous carbon dioxide in (\ref{eq:dbc1}) as follows.  Denoting the concentration of gaseous carbon dioxide CO$_{2_{(\mathrm{gas})}}$, in the concrete by $c_g$ with diffusion coefficient $D_{c_g}$, its exchange with the external atmospheric concentration $c_g^*$ at the exposed concrete surface is given by
\[
 \text{at} \, x=0 \hskip 1cm     -D_{c_g} \frac{\partial c_g}{\partial x} = H^*_g (c^*_g- c_g) ,
\]
where $H^*_g$ is the mass transfer constant.

Assuming an equilibrium balance $c=C^{H} c_g$ for the local exchange of gaseous and aqueous carbon dioxide, we may combine these expressions to give (\ref{eq:dbc1}), where
\[
   H^*=  \frac{H^*_g D_c}{D_{c_g}}, \hskip 1cm c^*=C^Hc^*_g.
\]
The dimensionless Henry constant $C^H$ is temperature dependent, a typical value being 0.82 at $20^oC$. We remark that the assumption of equilibrium balance
between gaseous and aqueous carbon dioxide means that $D_c$ should be taken as $D_{c_g}$, the governing equation (\ref{eq:dge1}) then being consistent with the models
of \cite{PVF,PVF2} when written in terms of $c_g$. More sophisticated models may relax this equilibrium assumption, allowing CO$_2$ in the gas and aqueous phases to be considered separately.

The formation of calcium carbonate (concentration $z(x,t)$) can be modelled using the rate equation
\be
      \frac{\partial z}{\partial t} =  R(c,h) ,
\ee
where its diffusivity is taken as being negligible. As such the amount of carbonate can be determined once (\ref{eq:dge1})--(\ref{eq:dic}) is solved together with specifying a suitable initial condition e.g. $z=0$ at $t=0$.

We non-dimensionalize as follows,
\begin{gather*}
x=L\bar{x}, \hskip 0.5cm  t=\frac{L^2 h^0}{D_{c}^0 c^*} \bar{t}, \hskip 0.5cm  c=c^*\bar{c}, \hskip 0.5cm h=h^0\bar{h},   \\
D_c=D_{c}^0\bar{D}_c, \hskip 0.5cm D_h=D_{h}^0\bar{D}_h,  \hskip 0.5cm R(c,h)=\theta r(\bar{c}, \bar{h}),
\end{gather*}
using $L$ as the characteristic length scale and $\theta$ a representative reaction rate scaling. We have taken $D_c^0$ and $D_h^0$ as the maximum values of the diffusion coefficients.  Dropping $\bar{}$ 's, we obtain
\begin{align}
\hskip -1cm \text{In} \, 0<x<1,t>0  \hskip 1cm  &\epsilon^2\left( \mu \frac{\partial c}{\partial t} - \frac{\partial}{\partial x}\left(D_c\frac{\partial c}{\partial x}\right)\right)=-r(c,h) , \label{govc}\\
& \epsilon^2\left(\frac{\partial h}{\partial t} - \frac{\delta^2}{\epsilon^2}\frac{\partial}{\partial x}\left(D_h\frac{\partial h}{\partial x}\right)\right)=-r(c,h) ; \label{govh}
\end{align}
with boundary conditions
\begin{align}
\text{on} \, x=0 \hskip 1cm & - D_c\frac{\partial c}{\partial x}=H(1 -  c), \; \frac{\partial h}{\partial x}=0, \\
\text{on} \, x=1  \hskip 1cm & \frac{\partial c}{\partial x} = 0, \, \frac{\partial h}{\partial x} = 0 ;
\end{align}
and initial conditions
\be
\text{at} \, t=0 \hskip 1cm  c= c_i(x), \hskip 0.5cm h=h_i(x) , \label{ic}
\ee
where
\begin{equation}
\epsilon^2= \frac{D_{c}^0c^*}{L^2\theta},\hskip 1cm \delta^2=\frac{D_{h}^0 h^0}{L^2\theta} , \hskip 1cm  \mu=\frac{c^*}{h^0},
 \hskip 1cm H=\frac{ L H^*}{D_{c}^0 }.
\label{eq:nondimpars}
\end{equation}
In the semi-infinite concrete case, $L$ is at our disposal and can be taken to be $D_c^0/H^*$ so that $H=1$.

The scaling $\theta$ is chosen so that $r(c,h)\le 1$ for $0\leq c \le 1$, $0\le h \le 1$. To be more precise, for reaction terms of the form
\begin{equation*}
R(c,h)=k c^ph^q
\end{equation*}
where $k$ is a constant reaction coefficient and $p$ and $q$ are positive constants, we have
\begin{equation}
r(c,h)=c^ph^q \label{reacterm}
\end{equation}
by choosing $\theta=k (c^{*})^p (h^{0})^q$. Referring to \cite{taylor} and  \cite{MPMB}, typical values of the dimensional parameters are
given in Table \ref{CCValues}. A representative length $L$ can be of the order of cms upto several metres i.e. $L=0.1 - 10$m depending upon the situation and geometry considered. The reaction rate corresponds to the common partial reaction orders $p=q=1$. However, other values of these, $0.9 \leq p,q \leq 1.5$, also seem appropriate.  These give the estimates showned in Table \ref{NDValues}.

The parameter range of relevance is thus
\be
       \epsilon \ll 1, \hskip 1cm \delta \leq O(\epsilon) , \hskip 1cm \mu \ll 1, \hskip 1cm H=O(1),
\ee
where we note that the ``relative transport parameter"
\be
 \frac{\delta^2}{\epsilon^2} = \frac{D_h^0 h^0}{D_c^0 c^* }
\ee
is typically small or at most order 1 \cite{PVF}. The CO$_2$ interfacial exchange parameter $H$ is estimated using the maximum values of the CO$_2$ diffusivities, which are expected to occur at the surface and for gas are denoted by $D_{c_g}^0$. Since $D_{c}^0=D_{c_g}^0$ we have that $H = L H_g^*/D_{c_g}^0$, the larger values of this parameter suggesting that the  interfacial exchanges are close to equilibrium.
The diffusion coefficients now have dependencies on the parameters $\epsilon$ and $\delta$ i.e. $D_c=D_c(c,h;\epsilon),D_h=D_h(c,h;\delta)$ being the general form. As far as this paper is concerned, we focus on the simpler dependency cases $D_c(h;\epsilon),D_h(c;\epsilon)$ the extensions being straightforward. For instance, the functional forms (\ref{eq:Di}) become
\be
    D_c = 1 - \exp \left( - \frac{\nu_1 \epsilon^2}{h^q} \right) , \hspace{0.5cm} D_h = 1 - \exp \left( - \frac{\nu_2 \delta^2}{c^p} \right),
\label{eq:DcDhArr}
\ee
where $\nu_1 \epsilon^2 = {\nu_c }/{h_0^q}, \nu_2 \delta^2= {\nu_h }/{c^{*p}}$ on taking $R_c=h^q,R_h=c^p$.

\begin{table}[t]
\centering
\begin{tabular}{lll}
\hline \vspace{-0.3cm}\\
\vspace{0.1cm}Parameter (units) & Value  (accelerated) & Value (natural) \\
\hline \\
\vspace{0.2cm}
$D_c^0=D_{c_g}^0$ \;(m$^2$s$^{-1}$) &  $(0.5 - 5)\times 10^{-8}$ & $(0.5 - 5)\times 10^{-8}$ \\
\vspace{0.2cm}
$D_h^0$  \;  (m$^2$s$^{-1}$)    & $10^{-13}$  &  $10^{-13}$ \\
\vspace{0.2cm}
$H_g^*$ \; (m s$^{-1}$)   & $1.16\times10^{-2}$  & $1.16\times10^{-2}$  \\
\vspace{0.2cm}
$c^*$ \; (mol m$^{-3}$) & $4.38$ & $2.71\times10^{-3}$  \\
\vspace{0.2cm}
$h^0$  \;  (mol m$^{-3}$) & $1.04\times10^3$ & $1.04\times10^3$    \\
\vspace{0.2cm}
$k$  \; (m$^3$mol$^{-1}$s$^{-1}$) & $1.74\times10^{-5}$ & $4.6\times10^{-5}$ \\
\hline
\end{tabular}
\caption{Typical parameter values for natural and accelerated\protect\footnotemark
carbonation obtained from \cite{MPMB}, \cite{SV2004}, \cite{PVF}, \cite{PVF2}. Concentrations are expressed
in units of mol m$^{-3}$ for consistency with the reaction rate (partial orders of the reaction being $p=q=1$); the molar masses of
44.01 g for CO$_2$ and 74 g for Ca(OH)$_2$ may be used to convert to units of g m$^{-3}$. }
\label{CCValues}
\end{table}
\footnotetext{ Natural environments normally contain $0.03-0.05\%$ CO$_2$, with the evolution of the carbonation depth being slow and taking many years (typically 5-10 years). An accelerated carbonation chamber, exposes concrete to $50\%$ CO$_2$ which dramatically reduces the time needed to perform experiments from years to a matter of days (typically upto 20 days). }

\begin{table}[t]
\centering
\begin{tabular}{cll}
\hline \vspace{-0.3cm}\\
\vspace{0.1cm} $\stackrel{\mbox{Nondimensional}}{\mbox{parameter}}$ & Value (accelerated) & Value (natural) \\
\hline\\
\vspace{0.2cm}
$\epsilon^2$ & $10^{-9} - 10^{-4}$ & $10^{-9} - 10^{-4}$ \\
\vspace{0.2cm}
$\delta^2$ & $10^{-11} - 10^{-7}$ & $10^{-10} - 10^{-5}$ \\
\vspace{0.2cm}
$\mu$ & $10^{-3}$ & $10^{-6}$ \\
\vspace{0.2cm}
$H$ & $10^4 - 10^7$ & $10^4 - 10^7$ \\
\hline
\end{tabular}
\caption{Estimated nondimensional parameters using values in Table \ref{CCValues}.} \label{NDValues}
\end{table}

\section{The sharp interface model derivation}\label{FBP_first}

The limit $\epsilon \rightarrow 0 $ will be considered; Ortoleva et al. \cite{Ortoleva} (cf. sect E, p. 1001) refer to this as the fast-aqueous-reaction asymptotics. This corresponds to the bulk reaction being very rapid and the reaction is expected therefore to be essentially in equilibrium ($c=0$ or $h=0$) except in a thin boundary layer, the reaction zone, where virtually all the carbonation occurs. In other words, almost all the reaction is confined to a very narrow layer, the structure of which will be sensitive to the behaviour of the diffusion coefficients. The location of the reaction zone will be denoted by $x=s(t;\epsilon)$. We denote the regions in which the reaction is in equilibrium as the outer regions, where outer 1 has $h=0$ and outer 2 has $c=0$. The details of the reaction zone depend crucially on the behaviour of the
diffusion coefficients, which we investigate in subsequent sections.  For generality in derivation, we keep $\mu=O(1)$ and consider its vanishing
limit later when seeking to solve the resulting sharp interface Stefan problems.

\subsection{Outer solutions}

Taking the limit $\epsilon \rightarrow 0$ and posing regular expansions
\begin{equation*}
c=c_0+ O(\epsilon^2) , \hskip 1cm  h=h_0+O(\epsilon^2) ,
\end{equation*}
we obtain two outer regions as below. The location of the reaction zone also requires expansion, the leading order term we
denote by $s(t)$, the higher order terms of which not impacting in the analysis to the orders of the calculations considered.

{\bf Outer1} $\quad 0<x<s(t)$ (carbon dioxide region)

Here we have $h_0=0$ (with in fact h=0 to all algebraic orders of $\epsilon$) and
\be
    \mu \frac{\partial c_0}{\partial t} - \frac{\partial}{\partial x} \left(D_{c_0}\frac{\partial c_0}{\partial x} \right)=0 ,
\label{eq:outer1c} \ee where $D_{c_0}=D_c(c_0,0;0)$. This region is essentially calcium hydroxide free, having been used up in the reaction to form
calcium carbonate.

{\bf Outer 2} $\quad s(t)<x<1$ (calcium hydroxide region)

In this region we have $c_0=0$ (and in fact $c=0$ to all algebraic orders in $\epsilon$) with
\be
   \frac{\partial h_0}{\partial t}- \frac{\delta^2}{\epsilon^2}\frac{\partial}{\partial x} \left( D_{h_0}\frac{\partial h_0}{\partial x}\right) =0 ,
\label{eq:outer2h} \ee with $D_{h_0}=D_h(0,h_0;0)$. This equation may simplify further depending upon the relative sizes of $\delta$ and $\epsilon$.
This region is almost pure calcium hydroxide, the presence of carbon dioxide being negligible.

\subsection{Interface conditions}

In practical terms it is the outer regions that are the ones of most significance; however, the interior layer must be analyzed to obtain the continuity conditions linking both outers. Without appealing to such an inner analysis, however, we can use \eqref{govc} and \eqref{govh} to obtain the statement
\begin{equation}
\frac{\partial}{\partial t}( \mu c - h)=\frac{\partial}{\partial x}\left(D_c\frac{\partial c}{\partial x} -  \frac{\delta^2}{\epsilon^2}D_h\frac{\partial h}{\partial x}\right).
\end{equation}
This may now be used to obtain the jump condition at $x=s(t)$,
\begin{equation}
\left[D_c\frac{\partial c}{\partial x} -  \frac{\delta^2}{\epsilon^2}D_h\frac{\partial h}{\partial x} + \dot{s}(\mu c - h)\right]^{s_+}_{s_-}=0,
\end{equation}
representing conservation of mass.
Using the outer solutions this yields,
\begin{equation}
\text{at} \, x=s(t), \quad -D_{c_0}\frac{\partial c_0(s^-,t) }{\partial x}- \frac{\delta^2}{\epsilon^2}D_{h_0}\frac{\partial h_0(s^+,t)}{\partial
x}=\dot{s}(\mu c_0(s^-,t)+ h_0(s^+,t)). \label{eq:jump1}
\end{equation}
This represents the Stefan type moving boundary condition for \eqref{eq:outer1c} and \eqref{eq:outer2h}. However, we still remain two conditions short in terms of specifying the moving boundary problem. These final conditions can only be found by undertaking an interior layer analysis of the reaction zone. They correspond to kinetic conditions and will take the general form
\begin{equation}
       c_0(s^-,t) = \Phi_1(\dot{s}), \hskip 1cm  h_0(s^+,t) = \Phi_2(\dot{s}), \label{eq:jump2}
\end{equation}
where the functions $\Phi_1$ and $\Phi_2$ are to be determined. To complete the sharp interface statement we require suitable initial conditions. These are parameter sensitive, the most usual situation being
\begin{align}
\text{at} \;\; t=0, \hskip 1cm  &c_0= c_i(x) \; \text{for $0<x<s(0)$}, \, h_0=h_i(x) \; \text{for $s(0)<x<1$} \notag\\
&\text{and $s(0)=s_0$},
\label{eq:icsharp}
\end{align}
with, more oftern than not,  the initial position of the reaction zone being taken at the concrete surface $s_0=0$. In certain parameter regimes, there is an initial carbonation stage in which the reaction zone remains in its initial location for a finite time $t_0$. In the this case we modify
(\ref{eq:icsharp}) to
\begin{align}
\text{at} \;\; t=t_0, \hskip 1cm  &c_0= C_i(x) \; \text{for $0<x<s(0)$}, \, h_0=H_i(x) \; \text{for $s(0)<x<1$}\notag\\ 
&\text{and $s(0)=s_0$},
\label{eq:icsharp2}
\end{align}
where $C_i$ and $H_i$ are the resulting concentration profiles at the end of this initial carbonation stage.

The moving boundary problem (\ref{eq:outer1c})--(\ref{eq:outer2h}), (\ref{eq:jump1})--(\ref{eq:jump2}) with (\ref{eq:icsharp}) or (\ref{eq:icsharp2})
is a two-phase problem in the regime $\delta=O(\epsilon)$. When $\delta \ll \epsilon$, it degenerates to a one-phase problem where $\Phi_2$ is no
longer needed.

\section{Reaction layer analysis: Slowly varying diffusivities}
\label{D0}

We start by considering the simplest case in which neither $D_c$ nor $D_h$ have any explicit dependance upon $\epsilon$. As such they remain order 1, which we refer to as slowly varying.
The reaction layer in these situations comprises a single region, which we describe here for the
parameter range $\delta \leq \epsilon$, larger values of $\delta$ not being physically relevant for the carbonation problem. These slowly varying diffusivities provide a useful base case against which to compare the effects of more rapidly varying diffusivities in later sections \ref{D1} and \ref{D2}.

\subsection{Asymptotic regimes}
\label{Regimes1}

\subsubsection{Case $\delta \leq \epsilon^{\frac{p+2}{p+1}}$}

Fullest balance in the governing equations is obtained when $\delta = \lambda \epsilon^{\frac{p+2}{p+1}}$ with $\lambda=O(1)$.
We introduce the scalings
\be
 x=s(t)+\epsilon^{\frac{2}{p+1}} \bar{x}, \hskip 1cm c=\epsilon^{\frac{2}{p+1}}\bar{c}, \hskip 1cm h=\bar{h},
\label{eq:c1scale}
\ee
which preserves the CO$_2$ flux from outer 1, to obtain
\begin{gather}
\epsilon^{\frac{4}{p+1}} \mu \frac{\partial \bar{c}}{\partial t} - \epsilon^{\frac{2}{p+1}} \mu \dot{s}\frac{\partial \bar{c}}{\partial \bar{x}}
- \frac{\partial}{\partial \bar{x}}\left( D_c \frac{\partial \bar{c}}{\partial \bar{x}}\right)= - \bar{c}^p \bar{h}^q, \label{eq:c1ge1} \\
\epsilon^{\frac{2}{p+1}}\frac{\partial \bar{h}}{\partial t} - \dot{s}\frac{\partial \bar{h}}{\partial \bar{x}} -
 \lambda^2 \frac{\partial}{\partial \bar{x}} \left( D_h\frac{\partial \bar{h} }{\partial \bar{x}} \right) = - \bar{c}^p \bar{h}^q, \label{eq:c1ge2}
\end{gather}
Posing
\be
\bar{c}=\bar{c}_0 + o(1), \hskip 1cm \bar{h}=\bar{h}_0 + o(1) ,
\label{eq:expan1}
\ee
with \be
    D_c(c,h) \sim D_c(0,\bar{h}_0), \hskip 1cm   D_h(c,h) \sim D_h(0,\bar{h}_0) ,
\ee
we obtain the leading order equations
\be
 \frac{\partial}{\partial \bar{x}}\left( D_c \frac{\partial \bar{c}_0}{\partial \bar{x}}\right)=  \bar{c}_0^p \bar{h}_0^q, \hskip 1cm
 \dot{s}\frac{\partial \bar{h}_0}{\partial \bar{x}} + \lambda^2
\frac{\partial}{\partial \bar{x}} \left( D_h\frac{\partial \bar{h}_0 }{\partial \bar{x}} \right) =  \bar{c}_0^p \bar{h}_0^q.  \label{eq:c1leadge}
\ee
The matching conditions for outer 1 and 2 are
\begin{eqnarray}
  && \text{as $\bar{x} \to -\infty$} \hskip 1cm
      D_c\frac{\partial \bar{c}_0}{\partial \bar{x}} \to D_c\frac{\partial {c}_0 (s^-,t)}{\partial {x} },
 \hskip 0.25cm \bar{h}_0 \to 0 ,  \label{eq:c1match1}\\
  && \text{as $\bar{x} \to +\infty$} \hskip 1cm \bar{c}_0 \to 0, \hskip 0.25cm
      \bar{h}_0 \to h_0(s^+,t) , \hskip 0.25cm \lambda^2 D_h\frac{\partial \bar{h}_0}{\partial \bar{x}} \to 0 ,  \label{eq:c1match2}
\end{eqnarray}
and necessarily $c_0(s^-,t)=0$. Consequently, we have
\be
       \Phi_1 = 0 ,
\label{eq:c1Phi}
\ee
whilst $\Phi_2$ is not required. The above analysis also holds for $\delta \ll \epsilon^{\frac{p+2}{p+1}}$ where we can simply take the limit $\lambda \to 0$.

\subsubsection{Case $\epsilon^{\frac{p+2}{p+1}} \ll \delta \ll \epsilon$}

The scalings in this case are given by
\be
 x=s(t)+ \delta \left(\frac{\epsilon}{\delta}\right)^p \bar{x}, \hskip 1cm c= \left(\frac{\delta}{\epsilon}\right)^2\bar{c}, \hskip 1cm h=\bar{h},
\label{eq:c2scale}
\ee
to obtain
\begin{gather}
 \delta^2 \left(\frac{\epsilon}{\delta}\right)^{2p} \mu \frac{\partial \bar{c}}{\partial t} -  \delta \left(\frac{\epsilon}{\delta}\right)^{p} \mu \dot{s}\frac{\partial \bar{c}}{\partial \bar{x}}
- \frac{\partial}{\partial \bar{x}}\left( D_c \frac{\partial \bar{c}}{\partial \bar{x}}\right)= - \bar{c}^p \bar{h}^q, \label{eq:c2ge1} \\
\epsilon^2 \left(\frac{\epsilon}{\delta}\right)^{2p} \frac{\partial \bar{h}}{\partial t}
  - \epsilon \left(\frac{\epsilon}{\delta}\right)^{p+1} \dot{s}\frac{\partial \bar{h}}{\partial \bar{x}}
 -  \frac{\partial}{\partial \bar{x}} \left( D_h\frac{\partial \bar{h} }{\partial \bar{x}} \right) = - \bar{c}^p \bar{h}^q. \label{eq:c2ge2}
\end{gather}
Posing
\be
\bar{c}\sim\bar{c}_0 + \epsilon \left(\frac{\epsilon}{\delta}\right)^{p+1} \bar{c}_1, \hskip 1cm \bar{h}\sim \bar{h}_0 + \epsilon \left(\frac{\epsilon}{\delta}\right)^{p+1} \bar{h}_1 ,
\label{eq:expan2}
\ee
we obtain the leading order equations
\begin{gather}
 \frac{\partial}{\partial \bar{x}}\left( D_c \frac{\partial \bar{c}_0}{\partial \bar{x}}\right)=  \bar{c}_0^p \bar{h}_0^q, \hskip 1cm
\frac{\partial}{\partial \bar{x}} \left( D_h\frac{\partial \bar{h}_0 }{\partial \bar{x}} \right) =  \bar{c}_0^p \bar{h}_0^q, \label{eq:c2leadge}
\end{gather}
whilst at first order we have (for brevity we only record the case in which both diffusivities constant, additional terms from the expansion of the diffusivities being present when they are concentration dependent)
\begin{gather}
 \frac{\partial}{\partial \bar{x}}\left( D_c \frac{\partial \bar{c}_1}{\partial \bar{x}}\right)=  p\bar{c}_0^{p-1} \bar{h}_0^q \bar{c}_1 +
q\bar{c}_0^{p} \bar{h}_0^{q-1} \bar{h}_1 , \hskip 1cm\\
\dot{s} \frac{\partial \bar{h}_0}{\partial \bar{x}} + \frac{\partial}{\partial \bar{x}} \left( D_h\frac{\partial \bar{h}_1 }{\partial \bar{x}} \right) =  p\bar{c}_0^{p-1} \bar{h}_0^q \bar{c}_1 +
q\bar{c}_0^{p} \bar{h}_0^{q-1} \bar{h}_1, \label{eq:c2firstge}
\end{gather}
together with the matching conditions for outer 1 and 2 being,
\begin{eqnarray}
  && \text{as $\bar{x} \to -\infty$} \hskip 0.8cm
       D_c\frac{\partial \bar{c}_0}{\partial \bar{x}} \to 0 ,
 \hskip 0.25cm
      D_c\frac{\partial \bar{c}_1}{\partial \bar{x}} \to D_c\frac{\partial {c}_0 (s^-,t)}{\partial {x} },\notag\\
  && \hspace{2.8cm} \bar{h}_0 \to 0 , \hskip 0.25cm \bar{h}_1 \to 0 , \label{eq:c2match1}\\
  && \text{as $\bar{x} \to +\infty$} \hskip 0.8cm \bar{c}_0 \to 0, \hskip 0.25cm  \bar{c}_1 \to 0, \hskip 0.25cm
      \bar{h}_0 \to h_0(s^+,t) , \hskip 0.25cm D_h\frac{\partial \bar{h}_0}{\partial \bar{x}} \to 0 ,\notag\\
  && \hspace{2.8cm} D_h\frac{\partial \bar{h}_1}{\partial \bar{x}} \to D_h\frac{\partial {h}_0 (s^+,t)}{\partial {x} } .    \label{eq:c2match2}
\end{eqnarray}
Necessarily $c_0(s,t)=0$ and thus (\ref{eq:c1Phi}) holds with again $\Phi_2$ not required. Here it is the first order terms which allow the fluxes to be matched with the outer regions.

\subsubsection{Case $\delta=O(\epsilon)$}
\label{D0c3}

In this parameter case, an altogether quite different scenario takes place. First there is an initial period in which the reaction zone remains at the surface. This period may be termed stage I carbonisation, using the terminology introduced for metal oxidation \cite{EK} where a similar situation
occurs. During this stage, the hydroxide at the surface is depleted, the details of which are described in Appendix \ref{appA}. This stage ends at a finite time
$t_0$ when the surface hydroxide concentration becomes small (i.e. zero at leading order). After this, stage II carbonation takes place where the reaction zone moves into the concrete, which we describe here.

Writing $\delta=\lambda \epsilon$ with $\lambda=O(1)$, the scalings are
\be
 x=s(t)+\epsilon^{\frac{2}{p+q+1}} \bar{x}, \hskip 1cm c=\epsilon^{\frac{2}{p+q+1}}\bar{c}, \hskip 1cm h=\epsilon^{\frac{2}{p+q+1}}\bar{h},
\ee
giving
\begin{gather}
\epsilon^{\frac{4}{p+q+1}} \mu \frac{\partial \bar{c}}{\partial t} - \epsilon^{\frac{2}{p+q+1}} \mu \dot{s}\frac{\partial \bar{c}}{\partial \bar{x}}
- \frac{\partial}{\partial \bar{x}}\left( D_c \frac{\partial \bar{c}}{\partial \bar{x}}\right)= - \bar{c}^p \bar{h}^q, \\
\epsilon^{\frac{4}{p+q+1}}\frac{\partial \bar{h}}{\partial t} - \epsilon^{\frac{2}{p+q+1}}\dot{s}\frac{\partial \bar{h}}{\partial \bar{x}} -
  \lambda^2 \frac{\partial}{\partial \bar{x}} \left( D_h\frac{\partial \bar{h} }{\partial \bar{x}} \right) = - \bar{c}^p \bar{h}^q.
\end{gather}
Posing (\ref{eq:expan1}), we obtain the leading order equations
\begin{gather}
 \frac{\partial}{\partial \bar{x}}\left( D_c \frac{\partial \bar{c}_0}{\partial \bar{x}}\right)=  \bar{c}_0^p \bar{h}_0^q, \hskip 1cm
  \lambda^2 \frac{\partial}{\partial \bar{x}}\left( D_h \frac{\partial \bar{h}_0}{\partial \bar{x}}\right) =  \bar{c}_0^p \bar{h}_0^q,
\end{gather}
where here $D_c(c,h)\sim D_c(0,0),D_h(c,h)\sim D_h(0,0)$ are constants. The matching conditions for outer 1 and 2 being,
\begin{eqnarray}
  && \text{as $\bar{x} \to -\infty$} \hskip 1cm
      D_c\frac{\partial \bar{c}_0}{\partial \bar{x}} \to D_c\frac{\partial {c}_0 (s^-,t)}{\partial {x} },
 \hskip 0.25cm \bar{h}_0 \to 0 ,  \\
  && \text{as $\bar{x} \to +\infty$} \hskip 1cm \bar{c}_0 \to 0, \hskip 0.25cm
           D_h\frac{\partial \bar{h}_0}{\partial \bar{x}} \to D_c\frac{\partial {h}_0 (s^+,t)}{\partial {x} }.
\end{eqnarray}
Now, we have $c_0(s^-,t)=0=h_0(s^+,t)$ and consequently,
\be
       \Phi_1 = 0 =\Phi_2.
\ee

\subsection{Numerical results} \label{NRS}

\begin{figure}[htbp]
\psfrag{0}{\tiny{$0$}}
\psfrag{0.2}{\hspace{-0.05cm}\tiny{$0.2$}}
\psfrag{0.4}{\hspace{-0.05cm}\tiny{$0.4$}}
\psfrag{0.6}{\hspace{-0.05cm}\tiny{$0.6$}}
\psfrag{0.8}{\hspace{-0.05cm}\tiny{$0.8$}}
\psfrag{1}{\hspace{-0.05cm}\tiny{$1$}}
\psfrag{t=0.1}{\tiny{$t=0.1$}}
\psfrag{t=0.2}{\tiny{$t=0.2$}}
\psfrag{t=0.3}{\tiny{$t=0.3$}}
\psfrag{t=0.4}{\tiny{$t=0.4$}}
\psfrag{A}{\tiny{Concentration}}
\psfrag{x}{\tiny{$x$}}
\centering
\subfigure[\small{$\delta=\epsilon^2$}]{
\label{d=e2S}
\includegraphics[scale=0.34]{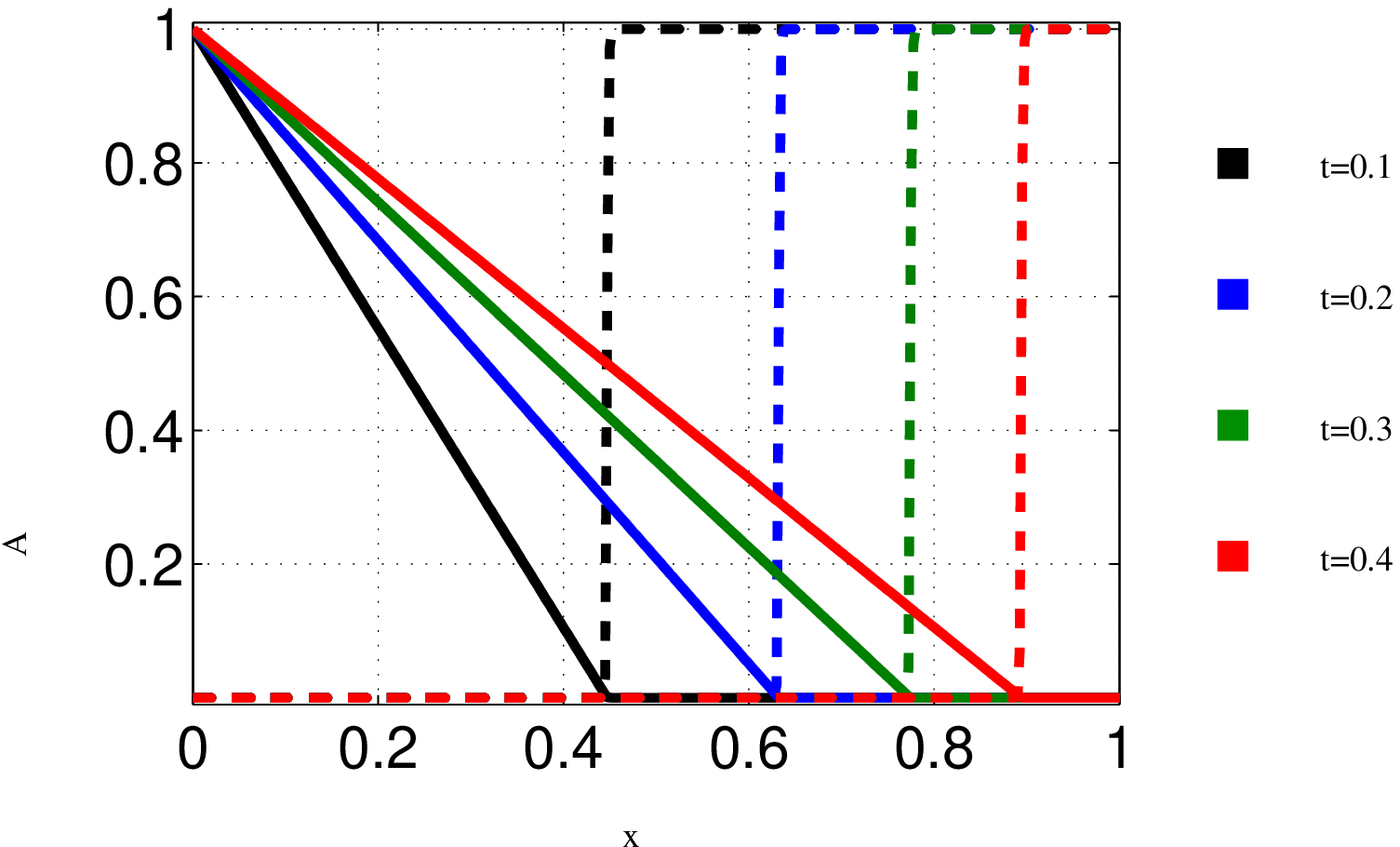}}
\hspace*{0.5cm}
\subfigure[\small{$\delta=\epsilon^{3/2}$}]{
\label{d=e1.5S}
\includegraphics[scale=0.34]{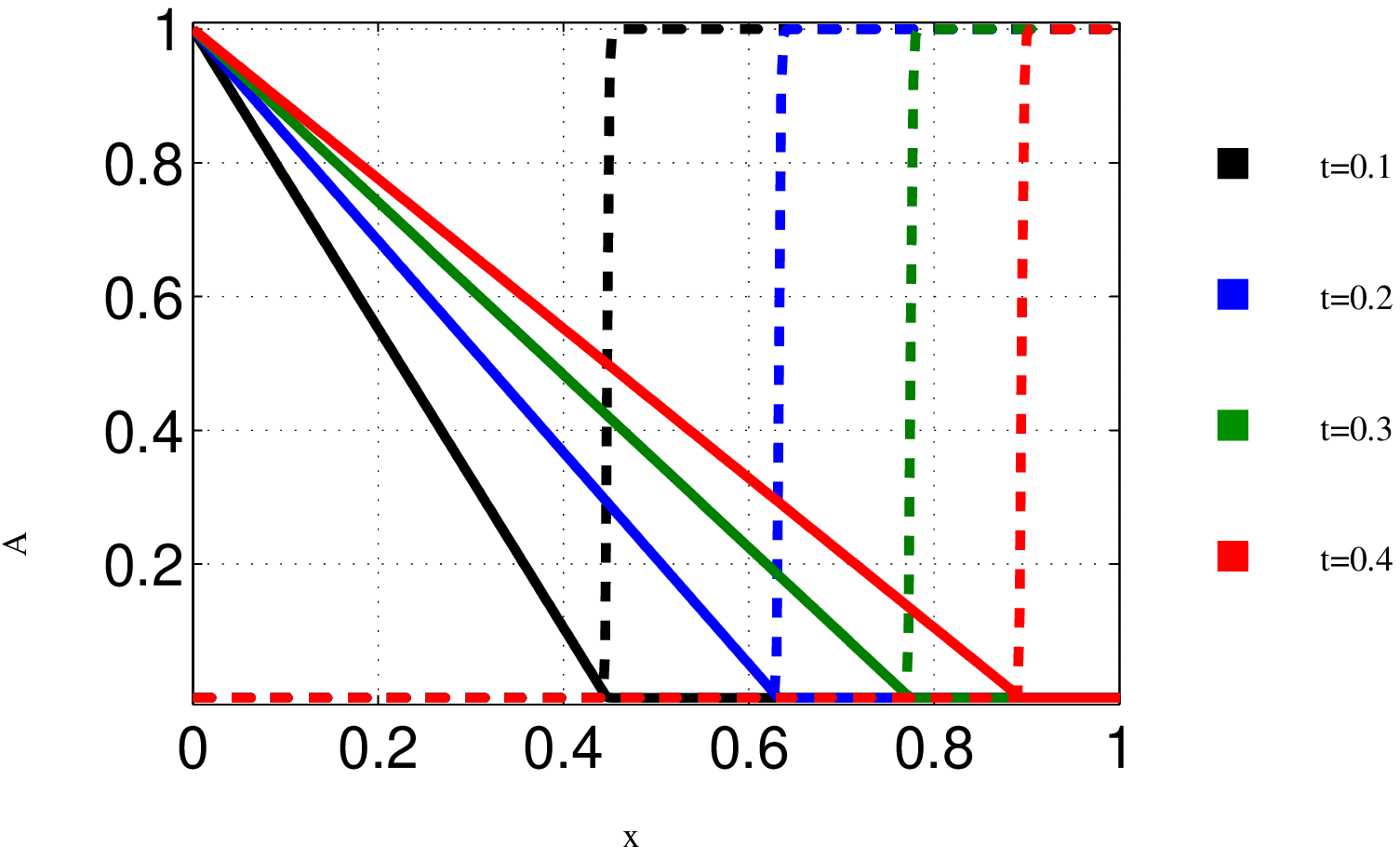}}
\subfigure[\small{$\delta=\epsilon^{1.25}$}]{ \label{d=e1.25S}
\includegraphics[scale=0.34]{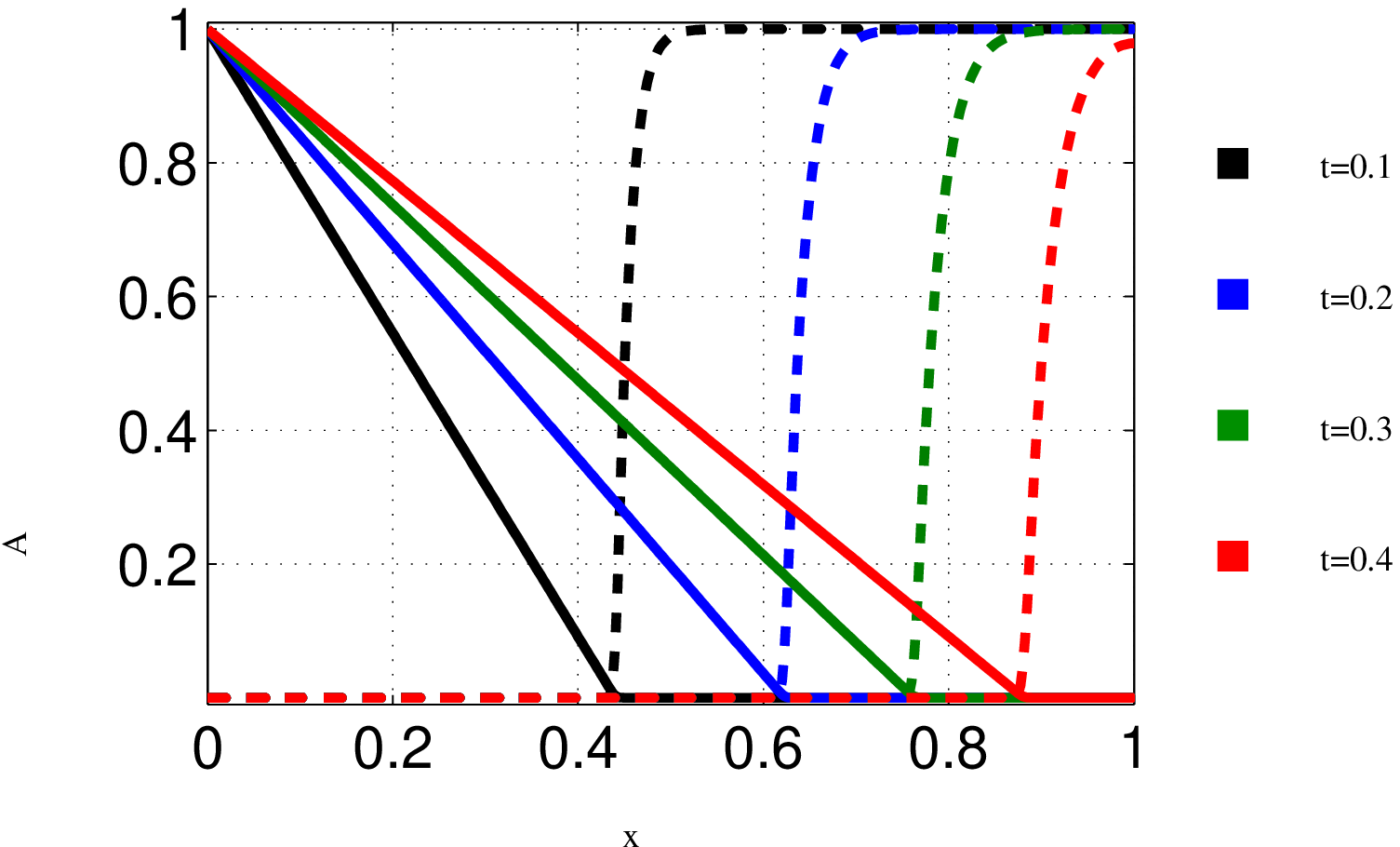}}
\hspace*{0.5cm}
\subfigure[\small{$\delta=\epsilon$}]{
\label{d=eS}
\includegraphics[scale=0.34]{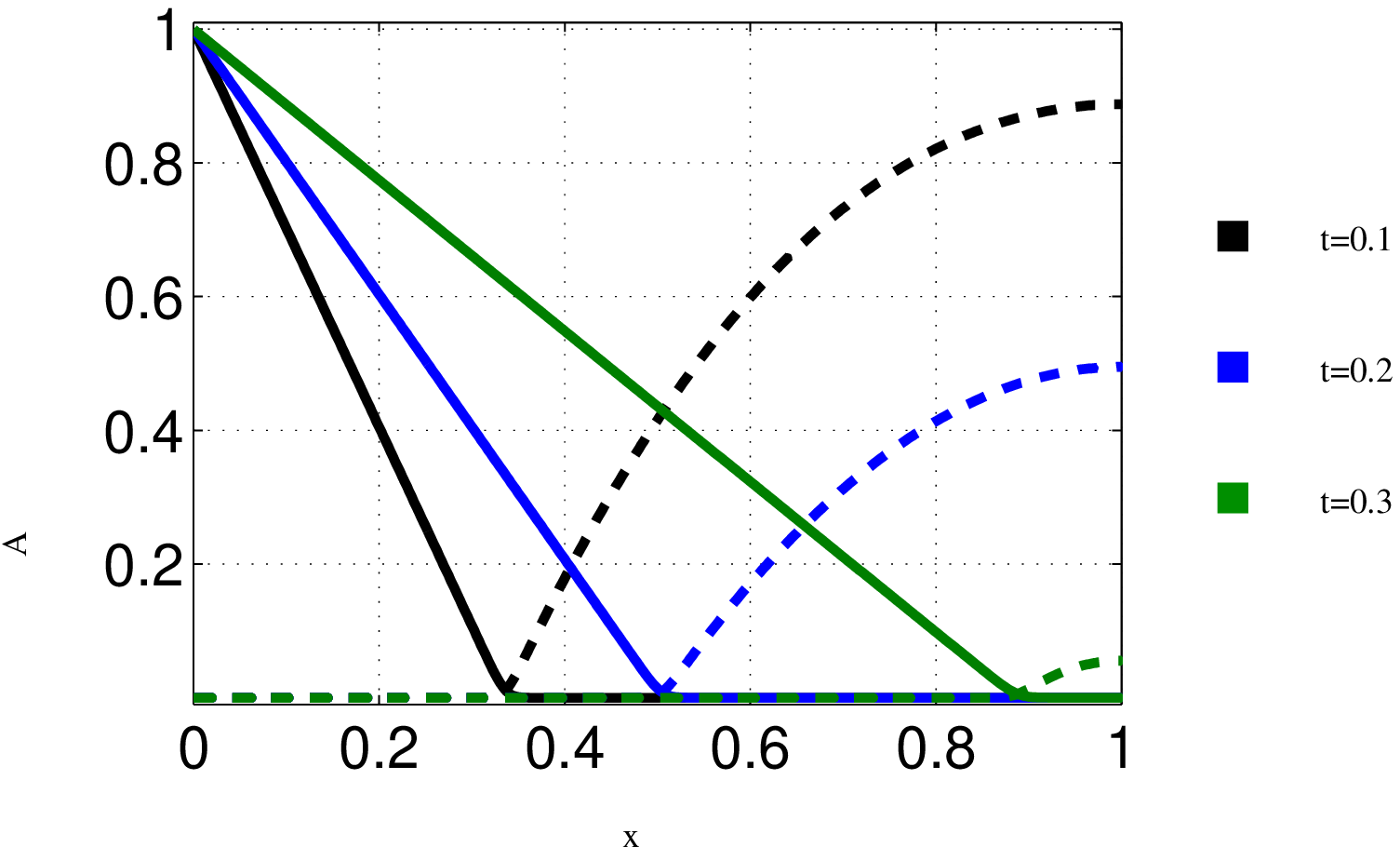}}
\caption{\small{Numerical results for slowly varying diffusivities. The parameter values are $\epsilon=10^{-3}$, $p=q=1$, $\mu=10^{-3}$ and $H=10^4$
and selected $\delta$ as stated in (a)--(d). The solid lines refer to the concentration of carbon dioxide while the dotted lines are the
concentration of calcium hydroxide at the shown times.}} \label{SlowGraph}
\end{figure}

Numerical solutions of the full model equations \eqref{govc}-\eqref{ic} are presented here for comparison with the asymptotics. The equations were
implemented in the finite element package COMSOL Multiphysics, using the general form PDE mode and a mesh of 400 quadratic Lagrange elements (1602
degrees of freedom). The time-dependent (Backward Difference Formula) solver had error tolerances abs tol$=10^{-6}$, rel tol$10^{-3}$. For all
numerical simulations we take initial data and parameter values
 \be
   h_i=1,c_i=0 \hspace{0.2cm} \mbox{for $0\leq x \leq 1$},
  \hspace{0.2cm} \mu=10^{-3}, \hspace{0.2cm} H=10^4, \hspace{0.2cm} p=q=1. \label{eq:numdata}
 \ee
Figure \ref{SlowGraph} shows numerical solutions for the case $D_c=D_h=1$, in the parameter regime $\epsilon=10^{-3}$. Four selected values of
$delta$ have been chosen, covering the asymptotic regimes presented in section \ref{Regimes1}. There is a clear marked change in behaviour between
the case $\delta<\epsilon$ shown in (a)--(c) and $\delta=\epsilon$ shown in (d). The hydroxide profile falls to zero at the reaction front in (d)
with noticeable diffusion ahead, in contrast to (a)--(c) where it is unity at the front and falls in zero within the reaction zone. Moreover, the
rates at which the fronts move vary, with a time delay apparent in (d) compared to (a)--(c). This supports the stage I carbonisation regime that is
postulated to take place when $\delta=\epsilon$, in which the reaction zone remains at the surface for finite time as the hydroxide concentration
falls. The reaction zone starts moving into the concrete once the hydroxide at the surface has been depleted.

\subsection{Sharp interface model summary}
\label{D0sum}

In the case $\delta \ll \epsilon$ we obtain the one-phase problem
\begin{eqnarray}
\lefteqn{\mbox{in} \; 0 < x < s(t),  t > 0}  \hspace{1.5in} &&   \mu \frac{\partial  c_0}
{\partial t} = \frac{\partial}{\partial x} \left( D_{c_0} \frac{\partial c_0}{\partial x} \right) ; \label{eq:s1ge} \\
\lefteqn{\mbox{on} \; x = 0} \hspace{1.5in} && -D_{c_0}\frac{\partial c_0}{\partial x}
= H(1 - c_0)  ; \\
\lefteqn{\mbox{on} \; x = s(t)} \hspace{1.5in} && c_0=0, \hskip 0.25cm  -D_{c_0}\frac{\partial {c_0}}{\partial x} =    \dot{s} h_i(s) ; \label{eq:s1mbc}\\
\lefteqn{\mbox{at} \; t = 0} \hspace{1.5in} && {c} = c_i \;  \text{for $0\leq x\leq s_i$}, \hskip 0.25cm s = s_i  ,
\label{eq:s1ic}
\end{eqnarray}
with $h_0 = h_i$ for $s(t)\leq x <1$.

In the case $\delta=\lambda \epsilon$ we obtain the two-phase problem
\begin{eqnarray}
\lefteqn{\mbox{in} \; 0 < x < s(t),  t > 0}  \hspace{1.5in} &&  \mu  \frac{\partial  c_0}
{\partial t} = \frac{\partial}{\partial x} \left( D_{c_0} \frac{\partial c_0}{\partial x} \right) ; \label{eq:s2ge1} \\
\lefteqn{\mbox{in} \; s(t) < x < 1,  t > 0}  \hspace{1.5in} &&   \frac{\partial  h_0}
{\partial t} = \lambda^2 \frac{\partial}{\partial x} \left( D_{h_0} \frac{\partial h_0}{\partial x} \right) ; \label{eq:s2ge2} \\
\lefteqn{\mbox{on} \; x = 0} \hspace{1.5in} && -D_{c_0}\frac{\partial c_0}{\partial x}
= H(1- c_0)  ; \\
\lefteqn{\mbox{on} \; x = 1} \hspace{1.5in} && D_{h_0}\frac{\partial h_0}{\partial x}
= 0  ; \\
\lefteqn{\mbox{on} \; x = s(t)} \hspace{1.5in} && c_0=0, \hskip 0.25cm h_0=0, \hskip 0.25cm  - D_{c_0}\frac{\partial c_0}{\partial x} =
       \lambda D_{h_0}\frac{\partial {h_0}}{\partial x}; \label{eq:s2mbc}\\
\lefteqn{\mbox{at} \; t = t_0} \hspace{1.5in} && c_0 = C_i \; \text{for $0\leq x\leq s_i$}, \hskip 0.25cm h_0=H_i \; \text{for $s_i\leq x\leq 1$}, \notag\\
 && {s} = s_i  .
\label{eq:s2ic}
\end{eqnarray}
Here $t_0$ is the end of the time at which the reaction zone remains at the outer surface and after which it begins to ingress into the concrete. $C_i,H_i$ denote the resulting concentration profiles at this time, which differ from their initial values $c_i,h_i$ respectively.

\section{Reaction layer analysis: A single rapidly varying diffusivity}
\label{D1}

Here we consider the carbon dioxide diffusivity to depend on $\epsilon$ as well as the hydroxide concentration, i.e. $D_c=D_c(c,h;\epsilon)$. The hydroxide remains slowly varying i.e. $D_h=O(1)$. The
properties required for this diffusivity are
\be
   D_c = \left\{  \begin{array}{ll}
                     O(\epsilon^2) & \text{for $h=O(1),h>0 ;$} \\
                     O(1) & \text{for $h=o(1),h>0.$}
                   \end{array}
     \right.
 \ee
Whilst the functional form is not important, for definiteness we may consider an hydroxide only dependent diffusitivity $D_c(h;\epsilon)$ in the Arrhenius form
\be
     D_c = 1 - \exp \left( -\frac{\nu_1 \epsilon^2}{h^{q}}\right),
\label{eq:DcArr}
\ee
where $\nu_1$ is a positive constant and $D_c(0;\epsilon)=1$. [A more general power of $h$ could be considered, with the below analysis only needing slight modification with the consideration of further terms in the expansions in the inner inner regions.]

\subsection{Asymptotic regimes}
\label{D1asymp}

\subsubsection{Case $\delta \leq \epsilon^2$}
\label{D1c1}

We write $\delta =\lambda \epsilon^2$ with $\lambda=O(1)$. The scalings for the inner inner region are
\be
 x=s(t)+\epsilon^{2} \bar{x}, \hskip 1cm c= \bar{c}, \hskip 1cm h=\bar{h}, \hskip 1cm D_c=\epsilon^2\bar{D}_c,
\label{eq:D1c31scale}
\ee
which give
\begin{gather}
\epsilon^{2} \mu \frac{\partial \bar{c}}{\partial t} - \mu \dot{s}\frac{\partial \bar{c}}{\partial \bar{x}}
- \frac{\partial}{\partial \bar{x}}\left( \bar{D}_c \frac{\partial \bar{c}}{\partial \bar{x}}\right)= - \bar{c}^p \bar{h}^q, \label{eq:D1c1ge1} \\
\epsilon^{2} \frac{\partial \bar{h}}{\partial t} - \dot{s}\frac{\partial \bar{h}}{\partial \bar{x}} -
 \lambda^2 \frac{\partial}{\partial \bar{x}} \left( D_h\frac{\partial \bar{h} }{\partial \bar{x}} \right) = - \bar{c}^p \bar{h}^q \label{eq:D1c1ge2}
\end{gather}
Posing (\ref{eq:expan1}) we obtain at leading order
\be
 \mu \dot{s}\frac{\partial \bar{c}_0}{\partial \bar{x}} + \frac{\partial}{\partial \bar{x}}\left( \bar{D}_c \frac{\partial \bar{c}_0}{\partial \bar{x}}\right)=  \bar{c}_0^p \bar{h}_0^q, \hskip 1cm
 \dot{s}\frac{\partial \bar{h}_0}{\partial \bar{x}} + \lambda^2
\frac{\partial}{\partial \bar{x}} \left( D_h\frac{\partial \bar{h}_0 }{\partial \bar{x}} \right) =  \bar{c}_0^p \bar{h}_0^q, \label{eq:D1c1leadge}
\ee
together with the matching conditions,
\begin{eqnarray}
  && \text{as $\bar{x} \to -\infty$} \hskip 1cm \bar{c}_0 \to c_0(s^-,t), \hskip 0.25cm
      \bar{D}_c\frac{\partial \bar{c}_0}{\partial \bar{x}} \to D_c\frac{\partial {c}_0 (s^-,t)}{\partial {x} },
 \hskip 0.25cm \bar{h}_0 \to 0 ,  \label{eq:D1c1match1}\\
  && \text{as $\bar{x} \to +\infty$} \hskip 1cm \bar{c}_0 \to 0, \hskip 0.25cm
      \bar{h}_0 \to h_0(s^+,t) , \hskip 0.25cm \lambda^2 D_h\frac{\partial \bar{h}_0}{\partial \bar{x}} \to 0 .  \label{eq:D1c1match2}
\end{eqnarray}

An additional region is required to facilitate the matching with outer 1 and explain (\ref{eq:D1c1match1}), due to the order of magnitude change in the carbon dioxide diffusivity. We denote this region as inner 1, the scalings for which are
\be
 x=s(t)+\epsilon^{ \frac{2}{q} } \hat{x}, \hskip 1cm c= \hat{c}, \hskip 1cm h= \epsilon^{\frac{2}{q}}\hat{h},
\label{eq:D1c1scale2}
\ee
giving
\begin{gather}
 \epsilon^{ \frac{4}{q} } \mu \frac{\partial \hat{c}}{\partial t} -  \epsilon^{ \frac{2}{q} } \mu \dot{s}\frac{\partial \hat{c}}{\partial \hat{x}}
- \frac{\partial}{\partial \hat{x}}\left( D_c \frac{\partial \hat{c}}{\partial \hat{x}}\right) = - \epsilon^{ \frac{4}{q} }\hat{c}^p \hat{h}^q, \label{eq:D1c1ige1} \\
 \epsilon^{ \frac{2}{q} } \frac{\partial \hat{h}}{\partial t}
  -  \dot{s}\frac{\partial \hat{h}}{\partial \hat{x}}
 - \lambda^2 \epsilon^{ \frac{2(q-1)}{q} } \frac{\partial}{\partial \hat{x}} \left( D_h\frac{\partial \hat{h} }{\partial \hat{x}} \right) = - \hat{c}^p \hat{h}^q. \label{eq:D1c1ige2}
\end{gather}
We restrict ourselves the parameter case $q>1$, the analysis needing slight modification for $0\leq q\leq 1$ which is described in Appendix \ref{appB}.
Posing
\be
\hat{c} \sim \hat{c}_0 + \epsilon^{ \frac{2}{q} } \hat{c}_1, \hskip 1cm \hat{h} \sim \hat{h}_0,
\label{eq:D1c1in1exp}
\ee
we obtain
\be
 \frac{\partial}{\partial \hat{x}}\left( D_c \frac{\partial \hat{c}_0}{\partial \hat{x}}\right) = 0 , \hskip 1cm
 \mu \dot{s}\frac{\partial \hat{c}_0}{\partial \hat{x}} +  \frac{\partial}{\partial \hat{x}}\left( D_c \frac{\partial \hat{c}_1}{\partial \hat{x}}\right) = 0 , \hskip 1cm
 \dot{s}\frac{\partial \hat{h}_0}{\partial \hat{x}} =  \hat{c}_0^p \hat{h}_0^q. \label{eq:D1c1ileadge}
\ee
After matching to outer 1 and the inner inner we have
\begin{gather}
    \hat{c}_0 = c_0(s^-,t) = \Phi_1, \hskip 0.25cm  D_c \frac{\partial \hat{c}_1}{\partial \hat{x}}= D_c \frac{\partial c_0(s^-,t)}{\partial x} = \lim_{\bar{x} \to -\infty} \bar{D}_c \frac{\partial \bar{c}_0}{\partial \bar{x}}  ,\notag\\ 
  \hat{h}_0 =  \left( - \frac{(q-1)}{\dot{s}}  c_0(s,t)^p \hat{x} \right)^{\frac{-1}{q-1}} .
\label{eq:D1c1insol}
\end{gather}

\subsubsection{Case $\epsilon^2 \ll \delta \ll \epsilon$}

\begin{figure}[htb]
\centering
\psfrag{x}{\footnotesize{$x$}}
\psfrag{s(t)}{\footnotesize{$s(t)$}}
\psfrag{Concentration}{\footnotesize{Concentration}}
\psfrag{0}{\footnotesize{$0$}}
\psfrag{1}{\footnotesize{$1$}}
\psfrag{O(y)}{\footnotesize{$O(\left(\frac{\delta}{\lambda\epsilon}\right)^2\theta)$}}
\psfrag{c=[CO_2]}{\footnotesize{$c=[\mathrm{C0}_2]$}}
\psfrag{h=[Ca(OH)_2]}{\footnotesize{$h=[\mathrm{Ca(OH)}_2]$}}
\psfrag{1}{\footnotesize{1}}
\psfrag{2}{\footnotesize{{\bf I1}}}
\psfrag{3}{\footnotesize{{\bf II}}}
\psfrag{4}{\footnotesize{{\bf I2}}}
\psfrag{O1}{\footnotesize{{\bf Outer 1}}}
\psfrag{O2}{\footnotesize{{\bf Outer 2}}}
\psfrag{O(z)}{\footnotesize{$O(\epsilon^{\frac{2}{q}})$}}
\psfrag{O(a)}{\footnotesize{$O(\left(\frac{\delta}{\lambda\epsilon}\right)^2)$}}
\psfrag{I1}{\footnotesize{I1 Inner 1}}
\psfrag{I2}{\footnotesize{I2 Inner 2}}
\psfrag{II}{\footnotesize{II Inner Inner}}
\includegraphics[scale=0.4]{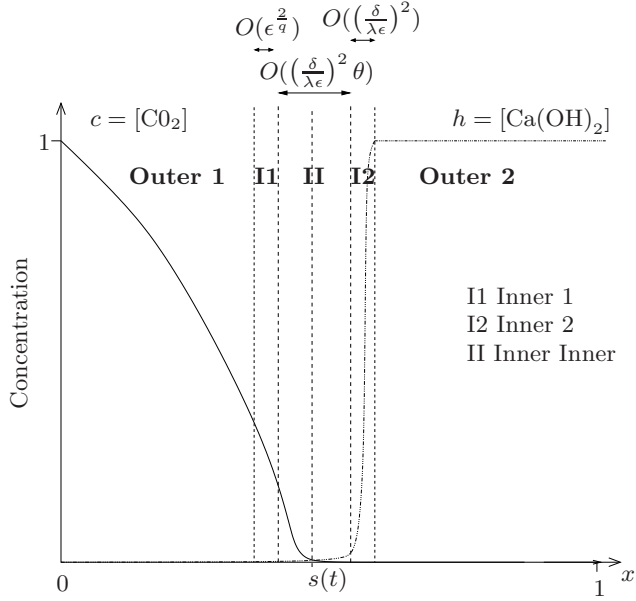}
\caption{\small Schematic summary of the asymptotic regions for a rapidly varying CO$_2$ diffusivity in the parameter case $\epsilon^2\ll\delta\ll\epsilon$.
The reaction layer lies between outer regions outer 1 ($0<x<s(t)$) and outer 2 ($s(t)<x<1$). It is composed of three regions: an inner inner region (II) with
$\theta=\left(\frac{\lambda\epsilon^2}{\delta}\right)^{\frac{2}{1+q}}$, together with inners 1 (I1) and 2 (I2). The inners 1 and 2 facilitate the matching of the inner inner region with the two outers. The case shown is for $q>1$ and $\delta\le\delta_{CR}$ with $\delta_{CR}=O(\epsilon^{\frac{1+q}{q}})$. Minor modifications are needed to inner 1 for the cases $q\le1$ and $\delta>\delta_{CR}$.}
\label{Sketch1}
\end{figure}

This case is more involved, where the reaction zone splits into three regions. In progressing from outer 2 to outer 1, we label the regions as
inner 2, inner inner and finally inner 1 as summarised in Figure \ref{Sketch1}.

The scalings for the inner 2 region are given by
\be
 x= s(t) +  \left( \frac{\delta}{\lambda \epsilon}\right)^2 \hat{x}, \hskip 1cm c= 0 , \hskip 1cm h= \hat{h},
\hskip 1cm D_c= \epsilon^2 \hat{D}_c ,
\label{eq:D1c2i2scale}
\ee
where $\lambda=O(1)$ is introduced for convenience and $c$ is actually exponentially small in $\epsilon$ if $p\geq 1$. The governing equation for the hydroxide at leading order in $\hat{x}>0$ is
\be
  \dot{s} \frac{\partial \hat{h}_0}{\partial \hat{x}}
      + \lambda^2 \frac{\partial}{\partial \hat{x}} \left( D_h\frac{\partial \hat{h}_0 }{\partial \hat{x}} \right) =  0 ,
      \label{eq:D1c2i2leadge}
\ee
together with the outer 2 matching conditions,
\begin{eqnarray}
    && \text{as $\hat{x} \to +\infty$} \hskip 1cm  \hat{h}_0  \to   h_0(s^+,t) , \hskip 0.25cm
     \frac{\partial \hat{h}_0}{ \partial \hat{x}} \to 0 .   \label{eq:D1c2i2match}
\end{eqnarray}
Thus
\[
           \lambda^2 D_h \frac{\partial \hat{h}_0}{ \partial \hat{x}} = \dot{s}(h_0(s^+,t)-\hat{h}),
\]
and for matching with the inner inner we have the behaviour
\begin{eqnarray}
    && \text{as $\hat{x} \to 0^+$} \hskip 1cm  \hat{h}_0  \to  0 , \hskip 0.25cm
    \lambda^2 D_h \frac{\partial \hat{h}_0}{ \partial \hat{x}} \to \dot{s} h_0(s^+,t) .   \label{eq:D1c2iimatch}
\end{eqnarray}
In the case $D_h$ a constant, we have the simple explicit solution
\[
      \hat{h}_0 = h_0(s^+,t) \left( 1- \exp\left( - \frac{\dot{s}}{\lambda^2 D_h} \hat{x}\right) \right) .
\]

For the inner inner region, we adopt the scalings
\be
 x= s(t) + \left( \frac{\delta}{\lambda \epsilon} \right)^2 \theta \bar{x}, \hskip 1cm c= \bar{c}, \hskip 1cm h= \theta \bar{h},
\hskip 1cm D_c= \frac{\epsilon^2}{\theta^q}  \bar{D}_c ,
\label{eq:D1c2iiscale}
\ee
where
\be
      \theta = \left( \frac{\lambda \epsilon^2}{\delta} \right)^{\frac{2}{1+q}} \ll 1.
\ee
The governing equations are now
\begin{gather}
  \left( \frac{\delta}{\lambda \epsilon} \right)^2 \theta \mu \frac{\partial \bar{c}}{\partial t} -  \mu \dot{s} \frac{\partial \bar{c}}{\partial \bar{x}}
- \frac{\partial}{\partial \bar{x}}\left( \bar{D}_c \frac{\partial \bar{c}}{\partial \bar{x}}\right) = -  \bar{c}^p \bar{h}^q, \\
  \left( \frac{\delta}{\lambda \epsilon} \right)^2 \theta^{2} \frac{\partial \bar{h}}{\partial t} -  \theta \dot{s} \frac{\partial \bar{h}}{\partial \bar{x}}
  - \lambda^2 \frac{\partial}{\partial \bar{x}} \left( D_h\frac{\partial \bar{h} }{\partial \bar{x}} \right) = - \bar{c}^p \bar{h}^q. \label{eq:D1c2iige}
\end{gather}
At leading order we have
\be
\mu \dot{s} \frac{\partial \bar{c}_0}{\partial \bar{x}}  +  \frac{\partial}{\partial \bar{x}}\left( \bar{D}_c \frac{\partial \bar{c}_0}{\partial \bar{x}}\right)=  \bar{c}_0^p \bar{h}_0^q, \hskip 1cm
 \lambda^2 \frac{\partial}{\partial \bar{x}} \left( D_h\frac{\partial \bar{h}_0 }{\partial \bar{x}} \right) = \bar{c}_0^p \bar{h}_0^q, \label{eq:D1c2iileadge}
\ee
together with the inner 2 matching conditions,
\be
   \text{as $\bar{x} \to +\infty $} \hskip 2cm \bar{c}_0 \to 0, \hskip 0.25cm
        \bar{D}_c \frac{\partial \bar{c}_0}{\partial \bar{x}} \to  0  , \hskip 0.25cm
         \bar{D}_h \frac{\partial \bar{h}_0}{\partial \bar{x}} \to \bar{D}_h \frac{\partial \hat{h}_0(0^+,t)}{\partial \hat{x}} ,
        \label{eq:D1c2iimatch2}
\ee
and the inner 1 matching behaviour
\be
   \text{as $\bar{x} \to -\infty$} \hskip 1.5cm \bar{c}_0 \to \hat{c}_0(0^-,t), \hskip 0.25cm
      \bar{D}_c \frac{\partial \bar{c}_0}{\partial \bar{x}} \to  \hat{D}_c \frac{\partial \hat{c}_0(0^-,t)}{\partial \hat{x}} ,
 \hskip 0.25cm \bar{h}_0 \to 0 .  \label{eq:D1c2iimatch1}
\ee

To allow matching between the inner inner and outer 1, an inner 1 region is required. The scalings are determined by the size of $\delta$ relative to
a critical value $\delta_{cr}=O(\epsilon^{\frac{(1+q)}{q}})$ and the value of $q$. We consider first the case $q>1$, for which $\epsilon^2 \ll \delta_{cr} \ll \epsilon$. For $\delta <\delta_{cr}$ the scalings (\ref{eq:D1c1scale2}) apply, as does the subsequent analysis culminating
in (\ref{eq:D1c1insol}). This also pertains when $\delta=\delta_{cr}$, with the modification of the diffusion term entering the leading order hydroxide equation. For $\delta > \delta_{cr}$, the scalings need adjusting to
\be
 x=s(t)+ \left( \frac{\delta}{\lambda \epsilon}\right) \epsilon^{ \frac{1}{q} } \hat{x}, \hskip 1cm c= \hat{c}, \hskip 1cm h= \epsilon^{\frac{2}{q}}\hat{h},
\label{eq:D1c2scale3}
\ee
which give
\begin{gather}
 \left( \frac{\delta}{\lambda \epsilon}\right)^2\epsilon^{ \frac{2}{q} } \mu \frac{\partial \hat{c}}{\partial t} - \left( \frac{\delta}{\lambda \epsilon}\right) \epsilon^{ \frac{1}{q} } \mu \dot{s}\frac{\partial \hat{c}}{\partial \hat{x}}
- \frac{\partial}{\partial \hat{x}}\left( D_c \frac{\partial \hat{c}}{\partial \hat{x}}\right) = - \left( \frac{\delta}{\lambda \epsilon}\right)^2 \epsilon^{ \frac{2}{q} }\hat{c}^p \hat{h}^q, \label{eq:D1c2i1ge1} \\
 \epsilon^{ \frac{2}{q} } \frac{\partial \hat{h}}{\partial t}
  -  \left( \frac{\epsilon}{\lambda \delta} \right) \epsilon^{ \frac{1}{q} } \dot{s}\frac{\partial \hat{h}}{\partial \hat{x}}
 -  \frac{\partial}{\partial \hat{x}} \left( D_h\frac{\partial \hat{h} }{\partial \hat{x}} \right) = - \hat{c}^p \hat{h}^q. \label{eq:D1c2i1ge2}
\end{gather}
Posing
\be
\hat{c} \sim \hat{c}_0 + \left( \frac{\delta}{\epsilon} \right) \epsilon^{ \frac{1}{q} } \hat{c}_1, \hskip 1cm \hat{h} \sim \hat{h}_0,
\label{eq:D1c2in1exp}
\ee
we obtain
\be
 \frac{\partial}{\partial \hat{x}}\left( D_c \frac{\partial \hat{c}_0}{\partial \hat{x}}\right) = 0 , \hskip 0.5cm
 \mu \dot{s}\frac{\partial \hat{c}_0}{\partial \hat{x}} +  \frac{\partial}{\partial \hat{x}}\left( D_c \frac{\partial \hat{c}_1}{\partial \hat{x}}\right) = 0 , \hskip 0.5cm
  \lambda^2 \frac{\partial}{\partial \hat{x}} \left( D_h \frac{\partial \hat{h}_0}{\partial \hat{x}} \right) =  \hat{c}_0^p \hat{h}_0^q. \label{eq:D1c2ileadge}
\ee

After matching to outer 1 and the inner inner, we again obtain (\ref{eq:D1c1insol})
with, for $q>1$,
\be
  \hat{h}_0 =
         \left( \frac{(q-1)^2c_0(s,t)^p}{2(1+q) D_h}  \hat{x}^2 \right)^{\frac{-1}{(q-1)}}
\label{eq:D1c2h0hat}
\ee
in the case when $D_h$ is independent of $\hat{x}$.

For the range $q\leq 1$, we note that $\delta_{cr} \leq \epsilon^2$ and so it is the regime just discussed that occurs with slight modification. Specifically, for $q=1$ this inner 1 region is that same size as the inner inner and so the spatial scaling in (\ref{eq:D1c2scale3}) is modified to
\be
   x = s(t) + \frac{\delta}{\lambda} (S(t) + \hat{x}) .
\ee
We again obtain (\ref{eq:D1c1ileadge})--(\ref{eq:D1c1insol}), but now with
\be
  \hat{h}_0 =  \exp \left( \sigma(t) \hat{x} \right) , \hskip 1cm S(t) \sim \frac{-2}{q(1+q) \sigma(t)} \log \left(
 \left(\frac{\lambda \epsilon}{\delta}\right)^{q} \frac{1}{\epsilon} \right),
\label{eq:D1c2i1sol3}
\ee
where
\[
\sigma(t) = \left( \frac{c_0(s^-,t)^p}{D_h } \right)^\frac{1}{2} .
\]

When $q<1$, the inner 1 region is narrower than the inner inner. We again have (\ref{eq:D1c1insol}) and (\ref{eq:D1c2h0hat}), where the latter
solution is taken for $\hat{x}>0$ and continued by zero for $\hat{x}<0$. The inner inner region is now restricted to $\bar{x}>0$ with the
inner 1 matching conditions (\ref{eq:D1c1insol}) occurring as $\bar{x} \to 0^+$.

\subsubsection{Case $\delta=O(\epsilon)$} \label{secd=eF1}

At the end of stage I carbonation (where the reaction zone remains at the outer surface), stage II carbonation takes place in which the reaction zone moves into the concrete.

In this case the reaction zone comprises two regions. We have an inner inner region that matches the order 1 flux of hydroxide from the outer 2 region as well as
allowing the CO$_2$ concentration to fall to $o(1)$. Again an inner 1 region is needed to accommodate the fall in CO$_2$ diffusivity, allowing matching between the inner inner and outer 1 regions. The details can be deduced from the previous subsection by setting $\delta=\lambda \epsilon$ with $\lambda=O(1)$. Consequently, for the inner inner region the scalings (\ref{eq:D1c2iiscale}) apply with $\theta=\epsilon^{\frac{2}{1+q}}$ so that again we obtain (\ref{eq:D1c2iileadge}) as the leading order equations. Since there is now no inner 2 region in this case, the matching conditions (\ref{eq:D1c2iimatch2}) now apply to outer 2, with $\hat{h}_0$ being replaced with $h_0$. The inner 1 matching conditions (\ref{eq:D1c2iimatch1}) remain the same as do the details for the inner 1 region as given by (\ref{eq:D1c2scale3})--(\ref{eq:D1c2i1sol3}).

\begin{figure}[htb]
\centering
\psfrag{x}{\footnotesize{$x$}}
\psfrag{s(t)}{\footnotesize{$s(t)$}}
\psfrag{Concentration}{\footnotesize{Concentration}}
\psfrag{0}{\footnotesize{$0$}}
\psfrag{1}{\footnotesize{$1$}}
\psfrag{O(y)}{\footnotesize{$O(\epsilon^{\frac{2}{q+1}})$}}
\psfrag{c=[CO_2]}{\footnotesize{$c=[\mathrm{C0}_2]$}}
\psfrag{h=[Ca(OH)_2]}{\footnotesize{$h=[\mathrm{Ca(OH)}_2]$}}
\psfrag{O1}{\footnotesize{{\bf Outer 1}}}
\psfrag{2}{\footnotesize{{\bf I1}}}
\psfrag{3}{\footnotesize{{\bf II}}}
\psfrag{O2}{\footnotesize{{\bf Outer 2}}}
\psfrag{O(a)}{\footnotesize{$O(\epsilon^{\frac{1}{q}})$}}
\psfrag{I1}{\footnotesize{I1 Inner 1}}
\psfrag{I2}{\footnotesize{II Inner Inner}}
\includegraphics[scale=0.4]{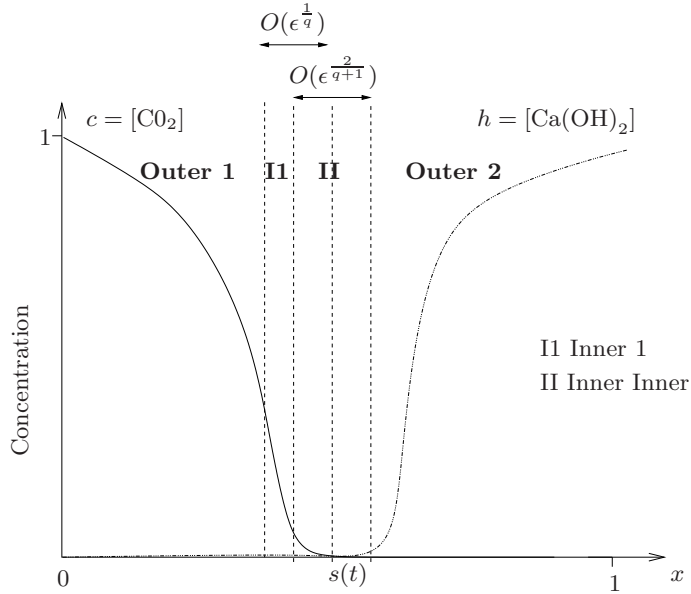}
\caption{\small A schematic of the asymptotic regions in the case $\delta=O(\epsilon)$ of a single rapidly varying CO$_2$ diffusivity.
The reaction zone is a two layer structure, comprised of an inner inner region of width $O \left(\epsilon^{\frac{2}{q+1}} \right)$, together with
an inner 1 of width $O\left(\epsilon^{\frac{1}{q}} \right)$. This latter transition region accommodates the change
in the rapidly varying diffusitivity, the situation $q>1$ being depicted.}
\label{Sketch3F}
\end{figure}

\subsection{Numerical results} \label{NRF1}

Here we present numerical simulations for a rapidly varying CO$_2$ diffusivity, using the scheme and data/parameter values specified in section
\ref{NRS}. We found it necessary to use the following regularisation for the diffusion coefficient \eqref{eq:DcArr},
\begin{equation}
 D_c=1-\mathrm{exp}\left(-\frac{\nu_1\epsilon^2}{(h^2+\theta_1^2)^{\frac{q}{2}}}\right), \label{reg}
\end{equation}
with $\theta_1=10^{-3}$. Although we don't present results for $p\neq 1,q\neq 1,$ we found the following regularisation of the reaction term
useful to avoid numerical difficulties,
\[
r={c}{(c^2+\theta_1^2)^{\frac{p-1}{2}}} {h}{(h^2+\theta_1^2)^{\frac{q-1}{2}}},
\]
particularly for $p<1$ or $q<1$.

\begin{figure}[htbp]
\psfrag{0}{\tiny{$0$}}
\psfrag{0.2}{\hspace{-0.05cm}\tiny{$0.2$}}
\psfrag{0.4}{\hspace{-0.05cm}\tiny{$0.4$}}
\psfrag{0.6}{\hspace{-0.05cm}\tiny{$0.6$}}
\psfrag{0.8}{\hspace{-0.05cm}\tiny{$0.8$}}
\psfrag{1}{\hspace{-0.05cm}\tiny{$1$}}
\psfrag{t=1.0}{\tiny{$t=1.0$}}
\psfrag{t=2.0}{\tiny{$t=2.0$}}
\psfrag{t=3.0}{\tiny{$t=3.0$}}
\psfrag{t=4.0}{\tiny{$t=4.0$}}
\psfrag{t=0.5}{\tiny{$t=0.5$}}
\psfrag{t=0.75}{\tiny{$t=0.75$}}
\psfrag{t=1.25}{\tiny{$t=1.25$}}
\psfrag{A}{\tiny{Concentration}}
\psfrag{x}{\tiny{$x$}}
\centering
\subfigure[\small{$\delta=\epsilon^2$}]{
\label{d=e2F1}
\includegraphics[scale=0.34]{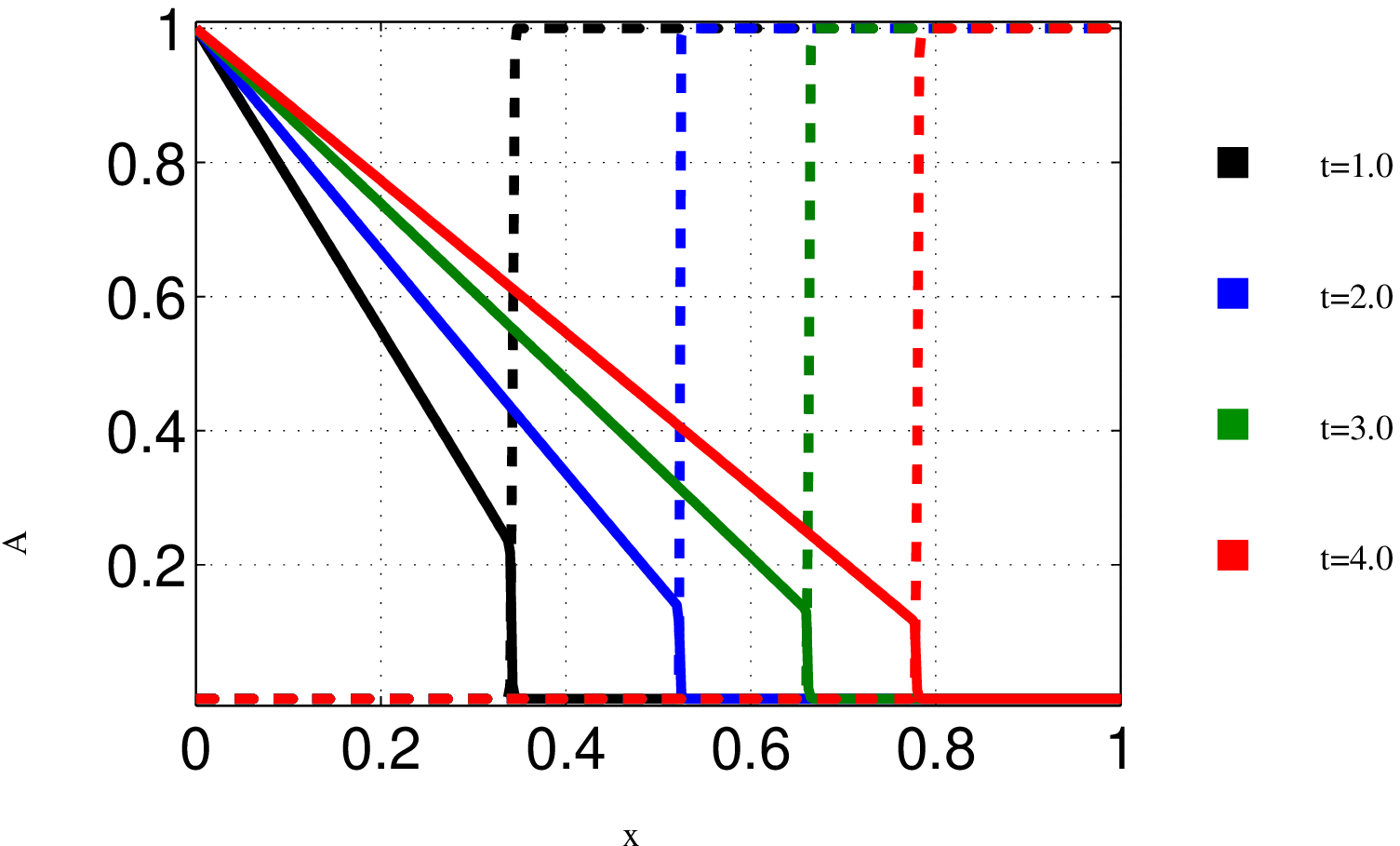}}
\hspace*{0.5cm}
\subfigure[\small{$\delta=\epsilon^{3/2}$}]{
\label{d=e1.5F1}
\includegraphics[scale=0.34]{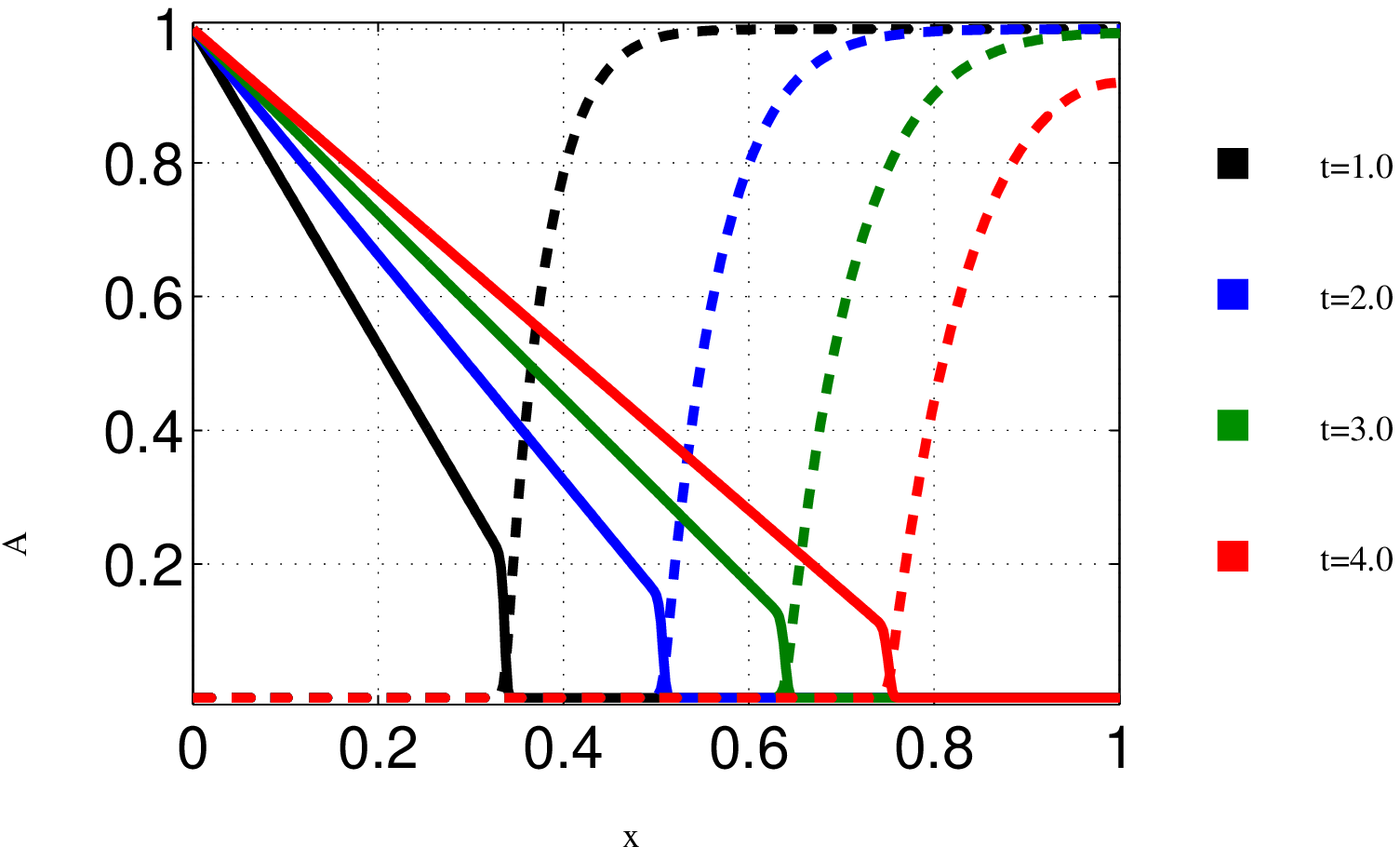}}
\subfigure[\small{$\delta=\epsilon$}]{
\label{d=eF1}
\includegraphics[scale=0.34]{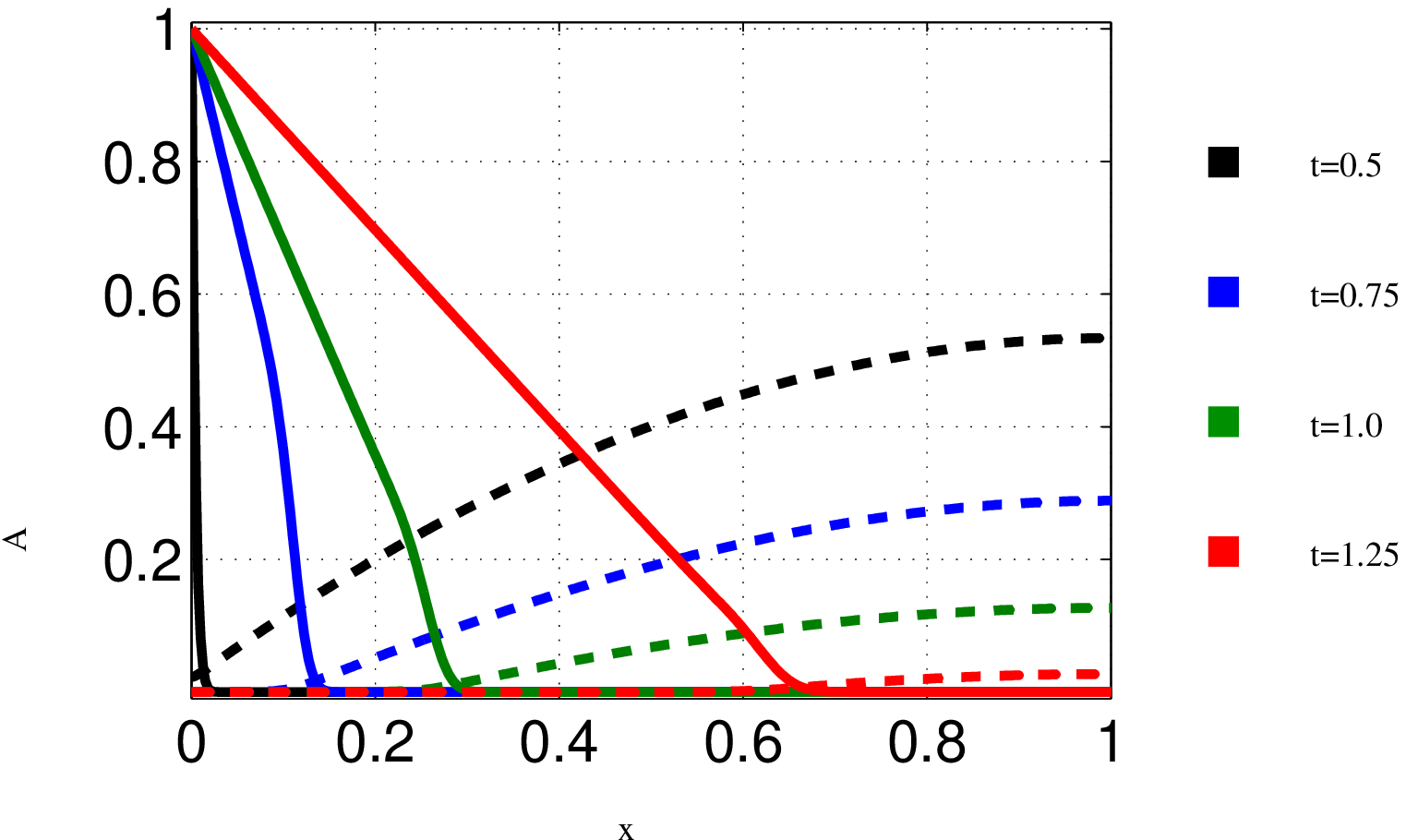}}
\caption{\small{Numerical results for one rapidly varying diffusion coefficient $D_c$ as defined in \eqref{eq:DcArr} for $\epsilon=10^{-2}$, $p=q=1$, $\mu=10^{-3}$ and $H=10^4$. The solid line refers to the concentration of carbon dioxide while the dotted line is the concentration of calcium hydroxide. }}
\label{Fast1Graph}
\end{figure}

Figure \ref{Fast1Graph} shows the numerical results for $\epsilon=10^{-2}$. Three selected cases of $\delta$ are shown covering the different
asymptotic regimes noted in the previous section. Noteworthy features are the $O(1)$ concentrations of CO$_2$ at the reaction zone in all cases. Also
the slower progress of the reaction zone compared to the corresponding situation in the slowly varying diffusivity case. Again the case
$\delta=O(\epsilon)$ shown in Figure \ref{d=eF1} gives different behaviour to the case $\delta \ll \epsilon$, with the hydroxide concentration being
small at the reaction zone front and stage I carbonisation taking place initially.

\subsection{Sharp interface model summary}
\label{D1sum}

In the case $\delta \leq \epsilon^2$, we now obtain the one-phase problem (\ref{eq:s1ge})--(\ref{eq:s1ic}) but with the more general moving boundary condition
\begin{eqnarray}
\lefteqn{\mbox{on} \; x = s(t)} \hskip 2cm && c_0=\Phi_1(\dot{s}), \hskip 0.25cm  -D_{c_0}\frac{\partial {c_0}}{\partial x} =   \dot{s}( \mu c_0+
h_i(s)), \label{eq:D1s1mbc}
\end{eqnarray}
in place of (\ref{eq:s1mbc}). We remark that $\Phi_1$ is determined by the solution of the inner inner problem
\be
 \mu \dot{s}\frac{\partial \bar{c}_0}{\partial \bar{x}} + \frac{\partial}{\partial \bar{x}}\left( \bar{D}_c \frac{\partial \bar{c}_0}{\partial \bar{x}}\right)=  \bar{c}_0^p \bar{h}_0^q, \hskip 1cm
 \bar{D}_c\frac{\partial \bar{c}_0 }{\partial \bar{x}}  - \lambda^2  D_h\frac{\partial \bar{h}_0 }{\partial \bar{x}} =  \dot{s} (\bar{h}_0 -h_i(s) - \mu \bar{c}_0), \label{eq:D1s1ii}
\ee
with
\begin{eqnarray}
  && \text{as $\bar{x} \to +\infty$} \hskip 1cm \bar{c}_0 \to 0, \hskip 0.25cm \bar{D}_c\frac{\partial \bar{c}_0 }{\partial \bar{x}} \to 0, \hskip 0.25cm
      \bar{h}_0 \to h_i(s) ,   \label{eq:c1iibc1}
\end{eqnarray}
that satisfies
\begin{eqnarray}
  && \text{as $\bar{x} \to -\infty$} \hskip 1cm
   \bar{c}_0 \to \Phi_1(\dot{s}) ,
 \hskip 0.25cm \bar{h}_0 \to 0, \hskip 0.25cm D_h\frac{\partial \bar{h}_0 }{\partial \bar{x}} \to 0.  \label{eq:c1iibc2}
\end{eqnarray}
For the functional form (\ref{eq:DcArr}) we have $\bar{D}_c = \nu_1/\bar{h}_0^{q}$.

For $\epsilon^2 \ll \delta \ll \epsilon$ we again have a one-phase model with moving boundary condition (\ref{eq:D1s1mbc}), but now $\Phi_1$ is determined by the solution of the inner inner 1 problem
\be
 \mu \dot{s}\frac{\partial \bar{c}_0}{\partial \bar{x}} + \frac{\partial}{\partial \bar{x}}\left( \bar{D}_c \frac{\partial \bar{c}_0}{\partial \bar{x}}\right)=  \bar{c}_0^p \bar{h}_0^q, \hskip 1cm
 \bar{D}_c\frac{\partial \bar{c}_0 }{\partial \bar{x}}  -  \lambda^2 D_h\frac{\partial \bar{h}_0 }{\partial \bar{x}} =  -\dot{s} (h_i(s) + \mu \bar{c}_0), \label{eq:D1s2ii}
\ee
with
\begin{eqnarray}
  && \text{as $\bar{x} \to +\infty$} \hskip 1cm \bar{c}_0 \to 0, \hskip 0.25cm \bar{D}_c\frac{\partial \bar{c}_0 }{\partial \bar{x}} \to 0, \hskip 0.25cm
    \lambda^2 D_h \frac{\partial  \bar{h}_0}{\partial \bar{x}} \to \dot{s} h_i(s)  ,   \label{eq:c2iibc1} \\
  && \text{as $\bar{x} \to -\infty$} \hskip 1cm
       \bar{c}_0 \to \Phi_1(\dot{s}) ,
 \hskip 0.25cm \bar{h}_0 \to 0, \hskip 0.25cm D_h\frac{\partial \bar{h}_0 }{\partial \bar{x}} \to 0 .  \label{eq:c2iibc2}
\end{eqnarray}

In the case $\delta =O(\epsilon)$, we now obtain the two-phase problem (\ref{eq:s2ge1})--(\ref{eq:s2ic}) but with the more general moving boundary condition
\begin{eqnarray}
\lefteqn{\mbox{on} \; x = s(t)} \hskip 2cm && c_0=\Phi_1(\dot{s}), \hskip 0.25cm  -D_{c_0}\frac{\partial {c_0}}{\partial x} - \lambda^2
D_{h_0}\frac{\partial {h_0}}{\partial x} =   \dot{s} \mu c_0, \label{eq:D1s2mbc}
\end{eqnarray}
in place of (\ref{eq:s2mbc}). We remark that the outer 2 solution determines $\Phi_2=0$, whilst $\Phi_1$ is determined by the solution of the inner inner problem
\be
\mu \dot{s} \frac{\partial \bar{c}_0}{\partial \bar{x}} +  \frac{\partial}{\partial \bar{x}}\left( \bar{D}_c \frac{\partial \bar{c}_0}{\partial \bar{x}}\right)=  \bar{c}_0^p \bar{h}_0^q, \hskip 1cm
 \lambda^2 \frac{\partial}{\partial \bar{x}} \left( D_h\frac{\partial \bar{h}_0 }{\partial \bar{x}} \right) = \bar{c}_0^p \bar{h}_0^q,
 \label{eq:D1s3ii}
\ee
with
\begin{eqnarray}
  && \text{as $\bar{x} \to +\infty$} \hskip 1cm \bar{c}_0 \to 0, \hskip 0.25cm \bar{D}_c\frac{\partial \bar{c}_0 }{\partial \bar{x}} \to 0, \hskip 0.25cm
      \bar{D}_h \frac{\partial \bar{h}_0 }{\partial \bar{x}} \to {D}_h \frac{\partial {h}_0(s^+,t) }{\partial {x}} ,   \label{eq:c3iibc1} \\
  && \text{as $\bar{x} \to -\infty$} \hskip 1cm
       \bar{c}_0 \to \Phi_1(\dot{s}) ,
 \hskip 0.25cm \bar{h}_0 \to 0 , \hskip 0.25cm \bar{D}_h \frac{\partial \bar{h}_0 }{\partial \bar{x}} \to 0.  \label{eq:c3iibc2}
\end{eqnarray}
This formulation holds for the parameter range $q\geq 1$. For $q<1$, the boundary condition (\ref{eq:c3iibc2}) is applied at $\bar{x}=0^+$ and the
problem considered on the half-line $\bar{x}>0$.

\section{Reaction layer analysis: Two rapidly varying diffusivities}
\label{D2}

We now consider the situation in which both the carbon dioxide and hydroxide diffusivities are rapidly varying. In addition to $D_c$ as defined in section (\ref{D1}) we consider an hydroxide diffusivity $D_h=D_h(c;\epsilon)$ with the behaviour
\be
   D_h = \left\{  \begin{array}{ll}
                     O(\delta^2) & \text{for $c=O(1),c>0 ;$} \\
                     O(1) & \text{for $c=o(1),c>0,$}
                   \end{array}
     \right.
 \ee
and $D_h(0;\epsilon)=1$. This maintains an $O(1)$ flux of hydroxide in the reaction zone. For definiteness, a specific functional form could be the  Arrhenius type,
\be
     D_h = 1 - \exp \left( -\frac{\nu_2 \delta^2}{c^p}\right),
\label{eq:DhArr}
\ee
where $\nu_2$ is a positive constant.

\subsection{Asymptotic regimes}

\subsubsection{Case $\delta=O(\epsilon)$}

Writing $\delta=\lambda \epsilon$ with $\lambda=O(1)$, the scalings for the inner inner region are
\be
 x=s(t)+\epsilon^{2} \bar{x}, \hskip 0.5cm c= \bar{c}, \hskip 0.5cm h=\bar{h}, \hskip 0.5cm D_c=\epsilon^2\bar{D}_c, \hskip 0.5cm
   D_h= \left(\frac{\delta}{\lambda} \right)^2 \bar{D}_h,
\label{eq:D2c1scale}
\ee
which give
\begin{gather}
\epsilon^{2} \mu \frac{\partial \bar{c}}{\partial t} - \mu \dot{s}\frac{\partial \bar{c}}{\partial \bar{x}}
- \frac{\partial}{\partial \bar{x}}\left( \bar{D}_c \frac{\partial \bar{c}}{\partial \bar{x}}\right)= - \bar{c}^p \bar{h}^q, \label{eq:D2ge1} \\
\epsilon^{2} \frac{\partial \bar{h}}{\partial t} - \dot{s}\frac{\partial \bar{h}}{\partial \bar{x}} -
  \left( \frac{\delta^2}{\lambda \epsilon^2}\right)^2 \frac{\partial}{\partial \bar{x}} \left( \bar{D}_h\frac{\partial \bar{h} }{\partial \bar{x}} \right) = - \bar{c}^p \bar{h}^q. \label{eq:D2ge2}
\end{gather}
Posing (\ref{eq:expan1}) we obtain at leading order
\be
 \mu \dot{s}\frac{\partial \bar{c}_0}{\partial \bar{x}} + \frac{\partial}{\partial \bar{x}}\left( \bar{D}_c \frac{\partial \bar{c}_0}{\partial \bar{x}}\right)=  \bar{c}_0^p \bar{h}_0^q, \hskip 1cm
 \dot{s}\frac{\partial \bar{h}_0}{\partial \bar{x}} +
\lambda^2 \frac{\partial}{\partial \bar{x}} \left( \bar{D}_h\frac{\partial \bar{h}_0 }{\partial \bar{x}} \right) =  \bar{c}_0^p \bar{h}_0^q, \label{eq:D2leadge}
\ee
together with the matching conditions,
\begin{eqnarray}
  && \text{as $\bar{x} \to -\infty$} \hskip 1cm
      \bar{D}_c\frac{\partial \bar{c}_0}{\partial \bar{x}} \to D_c\frac{\partial {c}_0 (s^-,t)}{\partial {x} },
 \hskip 0.25cm \bar{h}_0 \to 0 ,  \label{eq:D2match1}\\
  && \text{as $\bar{x} \to +\infty$} \hskip 1cm \bar{c}_0 \to 0, \hskip 0.25cm
      \bar{D}_h\frac{\partial \bar{h}_0}{\partial \bar{x}} \to D_h\frac{\partial {h}_0 (s^+,t)}{\partial {x} },   \label{eq:D2match2}
\end{eqnarray}

Again additional regions are required to facilitate the matching with outer 1 and 2, due to the order of magnitude change in the diffusivities. As in the single rapidly varying diffusivity case in section (\ref{D1asymp}) we require an inner 1 region, the scalings for which are
\be
 x=s(t)+\epsilon^{ \frac{2}{q} } \hat{x}, \hskip 1cm c= \hat{c}, \hskip 1cm h= \epsilon^{\frac{2}{q}}\hat{h}, \hskip 1cm D_h= \left( \frac{\delta}{\lambda} \right)^2 \hat{D}_h,
\label{eq:D2i1scale}
\ee
giving (\ref{eq:D1c1ige1})--(\ref{eq:D1c1ige2}), resulting in the same matching conditions (\ref{eq:D1c1insol}) (with care taken to distinguish the cases $0\leq q<1, q=1$ and $q>1$).

To facilitate the matching with outer 2, we consider an inner 2 region with the scalings
\be
 x=s(t)+ \epsilon^2 ( S(t;\epsilon)+\hat{x}), \hskip 1cm c= \epsilon^{\frac{2}{p}} \hat{c},
   \hskip 1cm h= \hat{h}, \hskip 1cm D_c= \epsilon^2\hat{D}_c,
\label{eq:D2c1scale2}
\ee
where $S(t;\epsilon) \gg 1$. Thus we have
\begin{gather}
 \epsilon^2  \mu \frac{\partial \hat{c}}{\partial t} -  \mu (\dot{s}+\epsilon^2 \dot{S} )\frac{\partial \hat{c}}{\partial \hat{x}}
-  \frac{\partial}{\partial \hat{x}}\left( \hat{D}_c \frac{\partial \hat{c}}{\partial \hat{x}}\right)
    = -  \epsilon^{ \frac{2(p-1)}{p} } \hat{c}^p \hat{h}^q, \label{eq:D2i2ge1} \\
 \epsilon^4  \frac{\partial \hat{h}}{\partial t}
  -  \epsilon^2  (\dot{s}+\epsilon^2\dot{S})\frac{\partial \hat{h}}{\partial \hat{x}}
 -  \lambda^2  \frac{\partial}{\partial \hat{x}} \left( D_h\frac{\partial \hat{h} }{\partial \hat{x}} \right)
  = - \epsilon^4  \hat{c}^p \hat{h}^q. \label{eq:D2i2ge2}
\end{gather}
We restrict ourselves first to the parameter range $p>1$. Posing
\be
\hat{c} \sim \hat{c}_0 , \hskip 1cm \hat{h} \sim \hat{h}_0 + \epsilon^2 \hat{h}_1,
\ee
we obtain
\be
 \frac{\partial}{\partial \hat{x}}\left( D_h \frac{\partial \hat{h}_0}{\partial \hat{x}}\right) = 0 , \hskip 0.5cm
 \frac{\partial}{\partial \hat{x}}\left( D_h \frac{\partial \hat{h}_1}{\partial \hat{x}}\right) = 0 , \hskip 0.5cm
 \mu \dot{s}\frac{\partial \hat{c}_0}{\partial \hat{x}}
    +  \frac{\partial}{\partial \hat{x}}\left( \hat{D}_c \frac{\partial \hat{c}_0}{\partial \hat{x}}\right)= 0 . \label{eq:D2i2leadge}
\ee
Thus we have
\begin{gather}
    \hat{h}_0 = h_0(s^+,t) = \Phi_2, \hskip 0.5cm  D_h \frac{\partial \hat{h}_1}{\partial \hat{x}}= D_h \frac{\partial h_0(s^+,t)}{\partial x} = \lim_{\bar{x} \to +\infty} \bar{D}_h \frac{\partial \bar{h}_0}{\partial \bar{x}}  , \notag\\
  \hat{c}_0 =  \text{exp} \left( - \frac{\hat{x}}{\dot{s}\hat{D}_c } \right) ,
\end{gather}
after matching to outer 2 and the inner inner regions, which also determines
\be
      S(t;\epsilon) \sim \frac{2\dot{s} \hat{D}_c}{p} \log(1/\epsilon) \hskip 1cm \text{as $\epsilon \to 0$}.
\ee
For $p=1$, slight modification of the above takes place with the leading order reaction term entering the equation for $\hat{c}_0$ in (\ref{eq:D2i2leadge}). For
$0< p<1$, the inner inner solution for $\bar{c}_0$ vanishes at $\bar{x}=S(t)$ say and we modify the spatial scaling in (\ref{eq:D2c1scale2}) to
\[
     x=s(t) + \epsilon^2 \left( S(t) + \epsilon^{\frac{(1-p)}{p}} \hat{x}\right).
\]
As a result the leading order equation for $\hat{c}$ becomes
\[
     \frac{\partial}{\partial \hat{x}}\left( \hat{D}_c \frac{\partial \hat{c}_0}{\partial \hat{x}}\right)
    = \hat{c}_0^p \hat{h}_0^q,
\]
with $\hat{D}_c=\nu_1/h_0(s^+,t)^q$ which posseses the explicit solution
\be
  \hat{c}_0 = \left\{ \begin{array}{ll}
         \left( \frac{(1-p)^2h_0(s^+,t)^p}{2(1+p) \hat{D}_c}  \hat{x}^2 \right)^{\frac{1}{1-p}} & \text{for $\hat{x}<0$}, \\
         0 &   \text{for $\hat{x}>0$}.
    \end{array} \right.
\ee

\subsubsection{Case $\delta \ll \epsilon$}

Only slight modification of the preceeding analysis is required for this case. The inner inner region remains the same with the scalings (\ref{eq:D2c1scale}), but
now (\ref{eq:D2leadge}) becomes
\be
 \mu \dot{s}\frac{\partial \bar{c}_0}{\partial \bar{x}} + \frac{\partial}{\partial \bar{x}}\left( \bar{D}_c \frac{\partial \bar{c}_0}{\partial \bar{x}}\right)=  \bar{c}_0^p \bar{h}_0^q, \hskip 1cm
 \dot{s}\frac{\partial \bar{h}_0}{\partial \bar{x}} =  \bar{c}_0^p \bar{h}_0^q, \label{eq:D2c2leadge}
\ee
with the second condition in (\ref{eq:D2match2}) replaced by
\be
  \text{as $\bar{x} \to +\infty$} \hskip 1.5cm  \bar{h}_0 \to h_0 (s^+,t) .
\ee

The inner 1 analysis with scalings (\ref{eq:D2i1scale}) remains the same for $q > 1$, whilst for $q=1$ the only change is omission of the hydroxide diffusion term which is subdominant. There is more of a change for $0<q<1$ where the spatial scaling and balance in the hydroxide equation now depend upon the size of $\delta$ relative to $\delta_{cr}=\epsilon^{\frac{(1+q)}{2q}}$. Specifically,

\noindent (i) for $\delta\leq \delta_{cr}$: $x=s(t) + \epsilon^2 \left( S(t) + \epsilon^{\frac{2(1-q)}{q}} \hat{x} \right) $  the dominant balance for the hydroxide equation being that in (\ref{eq:D1c1ileadge}) for $\delta<\delta_{cr}$ and  (\ref{eq:D1c1ileadgeb}) in Appendix \ref{appB} when $\delta=\delta_{cr}$;

\noindent (ii) for $\delta_{cr} < \delta < \epsilon$: $x=s(t) + \epsilon^2 \left( S(t) + \left(\frac{\delta}{\lambda \epsilon} \right)^2 \epsilon^{\frac{(1-q)}{q}} \hat{x} \right) $  the dominant balance for the hydroxide equation being (\ref{eq:D1c1ileadgec}) in Appendix \ref{appB}.

The inner 2 analysis remains unchanged for $\epsilon^2 \ll \delta \ll \epsilon$, where $\lambda=\delta/\epsilon \ll$ in (\ref{eq:D2i2ge2}). For $\delta=\epsilon^2$, the convection term balances the diffusion term in the hydroxide equation (\ref{eq:D2i2ge2}), although (\ref{eq:D2i2leadge}) remains unchanged. For $\delta \ll \epsilon^2$,
the hydroxide convection and diffusion terms balance in a thin region of size $\hat{x}=(\delta/\epsilon^2)^2$ centered at $S(t)$, which allows the required matching to outer 2.

\subsection{Numerical results} \label{NRF2}

We implement the scheme as in the previous numerical sections \ref{NRS} and \ref{NRF1}, using the same initial date and parameter values. An
analogous regularisation to (\ref{reg}) is adopted for the hydroxide diffusivity \eqref{eq:DhArr}.

\begin{figure}[htbp]
\psfrag{0}{\tiny{$0$}}
\psfrag{0.2}{\hspace{-0.05cm}\tiny$0.2$}
\psfrag{0.4}{\hspace{-0.05cm}\tiny$0.4$}
\psfrag{0.6}{\hspace{-0.05cm}\tiny$0.6$}
\psfrag{0.8}{\hspace{-0.05cm}\tiny$0.8$}
\psfrag{1}{\hspace{-0.05cm}\tiny$1$}
\psfrag{t=0.5}{\tiny{$t=1.0$}}
\psfrag{t=1.0}{\tiny{$t=2.0$}}
\psfrag{t=1.5}{\tiny{$t=3.0$}}
\psfrag{t=2.0}{\tiny{$t=4.0$}}
\psfrag{A}{\tiny{Concentration}}
\psfrag{x}{\tiny{$x$}}
\centering
\subfigure[\small{$\delta=\epsilon^2$}]{
\label{d=e2F2}
\includegraphics[scale=0.34]{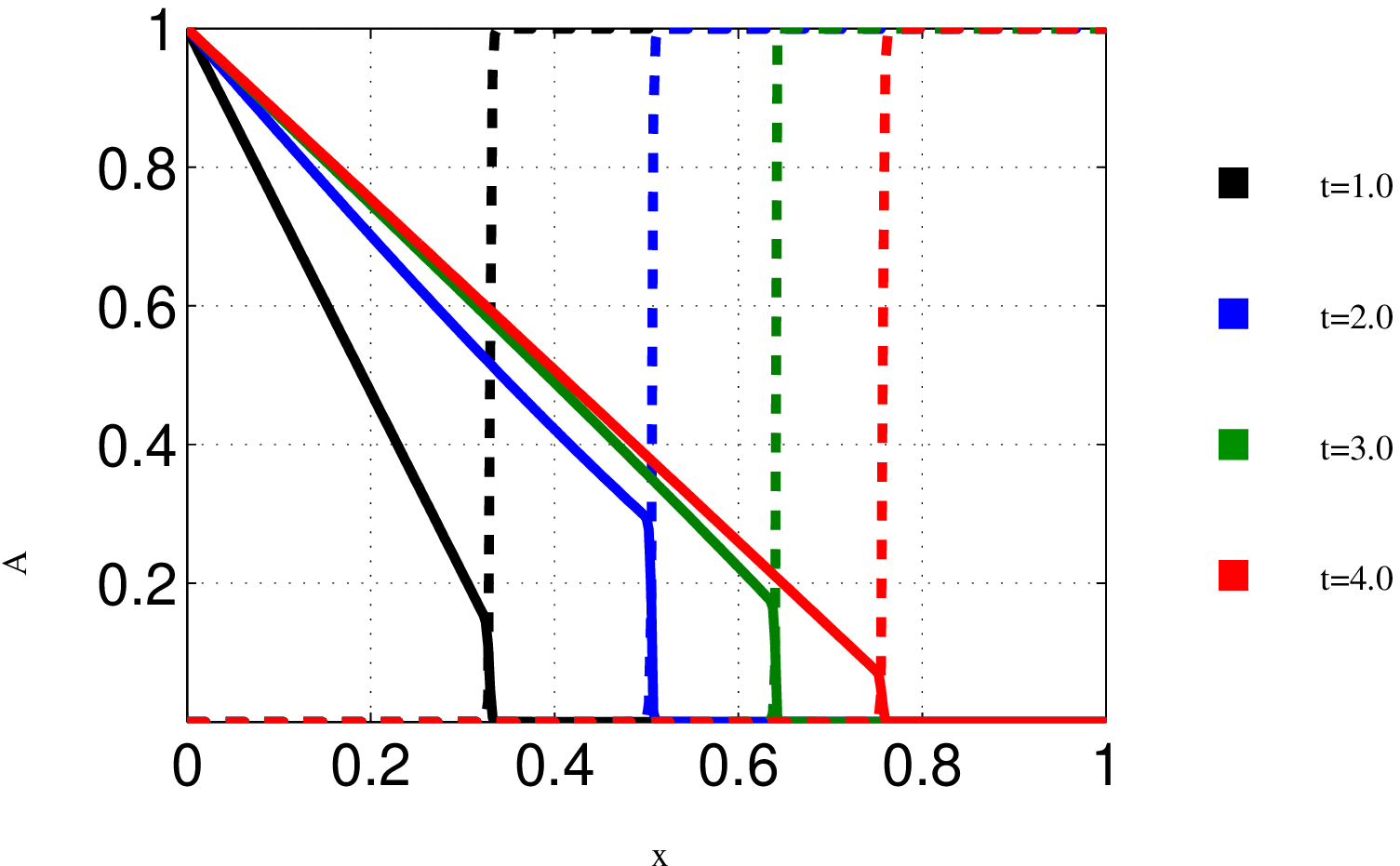}}
\hspace*{0.5cm}
\subfigure[\small{$\delta=\epsilon^{3/2}$}]{
\label{d=e1.5F2}
\includegraphics[scale=0.34]{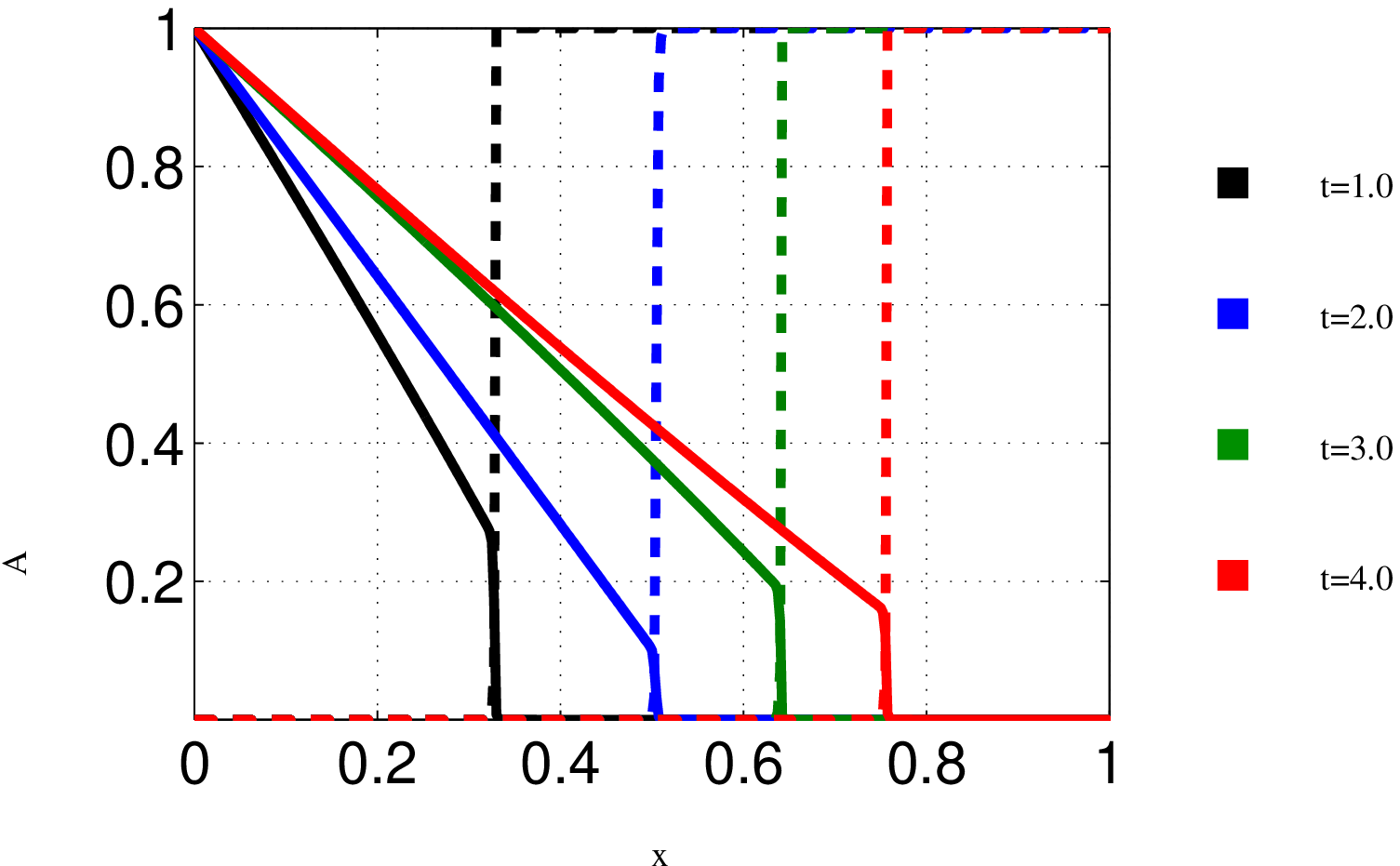}}
\subfigure[\small{$\delta=\epsilon$}]{ \label{d=e1F2}
\includegraphics[scale=0.34]{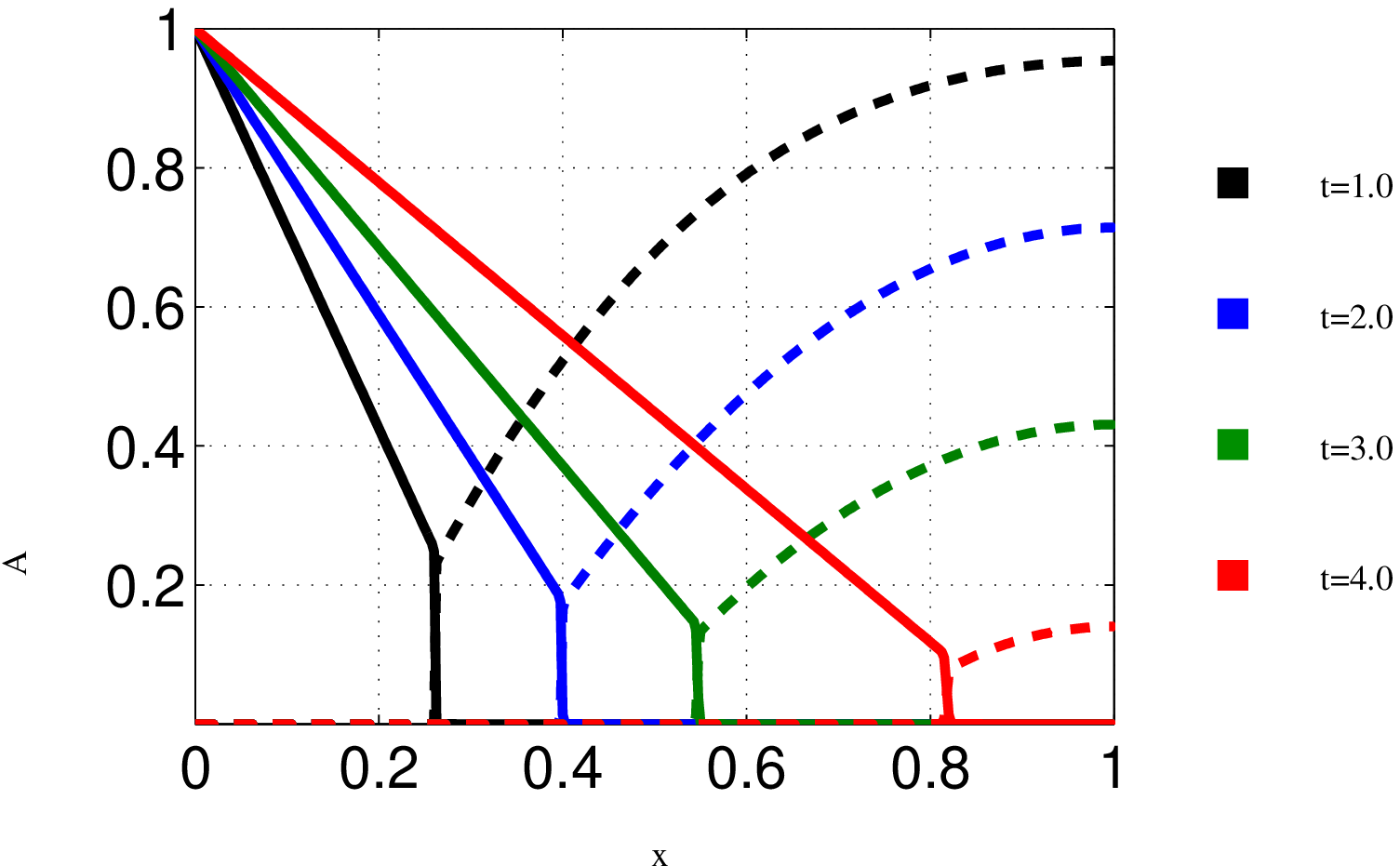}}
\caption{\small{Numerical results for two rapidly varying diffusivities. The parameter values are $\epsilon=10^{-2}$, $p=q=1$, $\mu=10^{-3}$ and $H=10^4$. The solid line refers to the concentration of carbon dioxide while the dotted line is the concentration of calcium hydroxide.}}
\label{Fast2Graph}
\end{figure}

Figure \ref{Fast2Graph} shows the numerical results for $\epsilon=10^{-2}$ and three selected values of  $\delta$.  Compared to the single rapidly
varying CO$_2$ diffusivity, the addition of a rapidly varying hydroxide diffusivity further slows down the progress of the reaction zone. The most
significant change occurs in the  $\delta=\epsilon$ regime, where now the hydroxide concentration is $O(1)$ in the reaction zone and no longer small
as for the slowly varying and single rapdily varying CO$_2$ cases in sections \ref{NRS} and \ref{NRF1}. Stage I carbonisation no longer takes place
in this case.

\subsection{Sharp interface model summary}
\label{D2sum}

In the case $\delta = \epsilon$, we obtain the two-phase problem (\ref{eq:s2ge1})--(\ref{eq:s2ic}) but with the more general moving boundary condition
\begin{eqnarray}
\lefteqn{\mbox{on} \; x = s(t)} \hskip 2cm && c_0=\Phi_1(\dot{s}), \hskip 0.25cm  h_0=\Phi_2(\dot{s}), \notag\\
  &&  -D_{c_0}\frac{\partial {c_0}}{\partial x} - \lambda^2 D_{h_0}\frac{\partial {h_0}}{\partial x} =   \dot{s}(\mu c_0+ h_0), \label{eq:D2mbc}
\end{eqnarray}
in place of (\ref{eq:s2mbc}). The solution of the inner inner problem
\be
 \mu \dot{s}\frac{\partial \bar{c}_0}{\partial \bar{x}} + \frac{\partial}{\partial \bar{x}}\left( \bar{D}_c \frac{\partial \bar{c}_0}{\partial \bar{x}}\right)=  \bar{c}_0^p \bar{h}_0^q, \hskip 1cm
   \dot{s}\frac{\partial \bar{h}_0}{\partial \bar{x}} + \frac{\partial}{\partial \bar{x}}\left( \bar{D}_h \frac{\partial \bar{h}_0}{\partial \bar{x}}\right)=  \bar{c}_0^p \bar{h}_0^q, \label{eq:D2c1ii}
\ee
with
\begin{eqnarray}
  && \text{as $\bar{x} \to +\infty$} \hskip 1cm \bar{c}_0 \to 0, \hskip 0.25cm \bar{D}_c\frac{\partial \bar{c}_0 }{\partial \bar{x}} \to 0,   \label{eq:D2iibc1} \\
 && \text{as $\bar{x} \to -\infty$} \hskip 1cm
      \bar{h}_0 \to 0, \hskip 0.25cm \bar{D}_h\frac{\partial \bar{h}_0 }{\partial \bar{x}} \to 0,   \label{eq:D2iibc2}
\end{eqnarray}
determines $\Phi_1,\Phi_2$ via
\be
   \Phi_1(\dot{s}) = \lim_{\bar{x} \to -\infty} \bar{c}_0 , \hskip 1cm \Phi_2(\dot{s}) = \lim_{\bar{x} \to +\infty} \bar{h}_0 .
\ee

In the case $\delta \ll \epsilon$, we have the one-phase problem (\ref{eq:s1ge})--(\ref{eq:s1ic}) with (\ref{eq:s1mbc}) replaced by
\begin{eqnarray}
\lefteqn{\mbox{on} \; x = s(t)} \hskip 2cm && c_0=\Phi_1(\dot{s}), \hskip 0.25cm  -D_{c_0}\frac{\partial {c_0}}{\partial x} =   \dot{s}(c_0+ h_i(s)),
\label{eq:D2s1mbc}
\end{eqnarray}
where $\Phi_1$ is determined by the solution of the inner inner problem
\be
 \mu \dot{s}\frac{\partial \bar{c}_0}{\partial \bar{x}} + \frac{\partial}{\partial \bar{x}}\left( \bar{D}_c \frac{\partial \bar{c}_0}{\partial \bar{x}}\right)=  \bar{c}_0^p \bar{h}_0^q, \hskip 1cm
 \bar{D}_c\frac{\partial \bar{c}_0 }{\partial \bar{x}}  -  \bar{D}_h\frac{\partial \bar{h}_0 }{\partial \bar{x}} =  \dot{s} (\bar{h}_0 -h_i(s) - \mu \bar{c}_0), \label{eq:D2c2ii}
\ee
with
\begin{eqnarray}
  && \text{as $\bar{x} \to +\infty$} \hskip 1cm \bar{c}_0 \to 0, \hskip 0.25cm \bar{D}_c\frac{\partial \bar{c}_0 }{\partial \bar{x}} \to 0, \hskip 0.25cm
      \bar{h}_0 \to h_i(s) ,   \label{eq:D2c2iibc1} \\
  && \text{as $\bar{x} \to -\infty$} \hskip 1cm
       \bar{c}_0 \to \Phi_1(\dot{s}) ,
 \hskip 0.25cm \bar{h}_0 \to 0 .  \label{eq:D2c2iibc2}
\end{eqnarray}

For the functional forms (\ref{eq:DcArr}) and (\ref{eq:DhArr}) we have $\bar{D}_c = \nu_1/\bar{h}_0^{q}, \bar{D}_h= \nu_2/\bar{c}_0^{p}$.

\section{The derived generalised Stefan problems} \label{SoltStefanProb}

\subsection{The one-phase problem}

In the case $\delta \ll \epsilon$ we obtain the one-phase problem
\begin{eqnarray}
\lefteqn{\mbox{in} \; 0 < x < s(t),  t > 0}  \hspace{1.5in} &&   \mu \frac{\partial  c_0}
{\partial t} = \frac{\partial}{\partial x} \left( D_{c_0} \frac{\partial c_0}{\partial x} \right) ; \label{eq:p1ge} \\
\lefteqn{\mbox{on} \; x = 0} \hspace{1.5in} && -D_{c_0}\frac{\partial c_0}{\partial x}
= H(1 - c_0)  ; \\
\lefteqn{\mbox{on} \; x = s(t)} \hspace{1.5in} && c_0=\Phi_1(\dot{s}), \hskip 0.25cm  -D_{c_0}\frac{\partial {c_0}}{\partial x} =   \dot{s}(\mu c_0+ h_i(s)), ; \label{eq:p1mbc}\\
\lefteqn{\mbox{at} \; t = 0} \hspace{1.5in} && {c} = c_i \;  \text{for $0\leq x\leq s_i$}, \hskip 0.25cm s = s_i  ,
\label{eq:p1ic}
\end{eqnarray}
with $h_0 = h_i$ for $s(t)\leq x <1$. For a slowly varying CO$_2$ diffusivity, the analysis of section \ref{D0} implies $\Phi_1=0$, whilst
for a rapidly varying CO$_2$ diffusivity $\Phi_1$ is given as the solution of the problem (\ref{eq:D1s1ii})--(\ref{eq:c1iibc2}) or (\ref{eq:D1s2ii})--(\ref{eq:c2iibc2}) depending upon the size of $\delta$ relative to $\epsilon^2$, or (\ref{eq:D2c2ii})--(\ref{eq:D2c2iibc2}) if the
hydroxide diffusivity varies rapidly as well.

The parameter $1/\mu$ represents the Stefan number, the time scaling $t=\mu \tau$ seeing it arise in the Stefan condition in (\ref{eq:p1mbc}) for its
more common occurrence. The asymptotics of this problem for power law forms of $\Phi_1=\dot{s}^n$ are described in \cite{EK2,EK3}. Relevant to the
concrete situation is the large Stefan number limit $\mu \to 0$. In this limit, the problem becomes quasi-steady due to the now disparate time-scales
of diffusion and reaction. This quasi-steady problem holds after an initial transient regime $t=O(\mu)$ in which the interface is stationary at
leading order. An explicit solution for the quasi-steady problem $t>O(\mu)$ when $D_{c_0}\equiv 1$ is \be
     c_0= \Phi_1 + \frac{(1-\Phi_1)(s-x)}{\left( \frac{1}{H} + s \right)},
\label{eq:c0qs}
\ee
with the interface $s(t)$ determined by the solution of
\be
        h_i(s)\left( \frac{1}{H} + s \right) \dot{s} + \Phi_1(\dot{s}) = 1, \hspace{1cm} s(0) = s_i .
\ee

\subsection{The two-phase problem}

In the case $\delta=\lambda \epsilon$ we obtain the two-phase problem
\begin{eqnarray}
\lefteqn{\mbox{in} \; 0 < x < s(t),  t > 0}  \hspace{1.5in} &&  \mu  \frac{\partial  c_0}
{\partial t} = \frac{\partial}{\partial x} \left( D_{c_0} \frac{\partial c_0}{\partial x} \right) ; \label{eq:p2ge1} \\
\lefteqn{\mbox{in} \; s(t) < x < 1,  t > 0}  \hspace{1.5in} &&   \frac{\partial  h_0}
{\partial t} = \lambda^2 \frac{\partial}{\partial x} \left( D_{h_0} \frac{\partial h_0}{\partial x} \right) ; \label{eq:p2ge2} \\
\lefteqn{\mbox{on} \; x = 0} \hspace{1.5in} && -D_{c_0}\frac{\partial c_0}{\partial x}
= H(1- c_0)  ; \\
\lefteqn{\mbox{on} \; x = 1} \hspace{1.5in} && D_{h_0}\frac{\partial h_0}{\partial x}
= 0  ; \\
\lefteqn{\mbox{on} \; x = s(t)} \hspace{1.5in} && c_0=\Phi_1(\dot{s}), \hskip 0.25cm h_0=\Phi_2(\dot{s}), \notag\\
    &&- D_{c_0}\frac{\partial c_0}{\partial x} -\lambda^2 D_{h_0}\frac{\partial {h_0}}{\partial x} = \dot{s}\left( \mu c_0 + h_0\right); \label{eq:p2mbc}\\
\lefteqn{\mbox{at} \; t = t_0} \hspace{1.5in} && c_0 = C_i \; \text{for $0\leq x\leq s_i$}, \hskip 0.25cm h_0=H_i \; \text{for $s_i\leq x\leq 1$},\notag\\
&&\hskip0.25cm  {s} = s_i  . \label{eq:p2ic}
\end{eqnarray}
Here $t_0$ is the end of the time at which the reaction zone remains at the outer surface and after which it begins to ingress into the concrete. $C_i,H_i$ denote the resulting concentration profiles at this time, which differ from their initial values $c_i,h_i$ respectively.
For slowly varying CO$_2$ and Ca(OH)$_2$ diffusivities, the analysis of section \ref{D0} implies $\Phi_1=0=\Phi_2$. For a single
rapidly varying CO$_2$ diffusivity $\Phi_1$ is given as the solution of the problem (\ref{eq:D1s3ii})--(\ref{eq:c3iibc2}) with $\Phi_2=0$. In both these situations there is a stage I carbonisation period where the Ca(OH)$_2$ concentration in the initial reaction zone at the surface falls to zero. When both diffusivities are rapidly varying then  $\Phi_1,\Phi_2$ are determined by (\ref{eq:D2c1ii})--(\ref{eq:D2iibc2}), there not being a stage I carbonisation period in this case.

In the quasi-steady CO$_2$ limit $\mu \to 0$, we again have solution (\ref{eq:c0qs}) in the case $D_{c_0}\equiv 1$. This gives the non-standard
Stefan problem for the Ca(OH)$_2$ as
\begin{eqnarray}
\lefteqn{\mbox{in} \; s(t) < x < 1,  t > 0}  \hspace{1.5in} &&   \frac{\partial  h_0}
{\partial t} = \lambda^2 \frac{\partial}{\partial x} \left( D_{h_0} \frac{\partial h_0}{\partial x} \right) ; \label{eq:p2bge2b} \\
\lefteqn{\mbox{on} \; x = 1} \hspace{1.5in} && D_{h_0}\frac{\partial h_0}{\partial x}
= 0  ; \\
\lefteqn{\mbox{on} \; x = s(t)} \hspace{1.5in} && \left\{\begin{array}{rcl}h_0&=&\Phi_2(\dot{s}),\\
    \lambda^2 D_{h_0}\frac{\partial {h_0}}{\partial x} &=& \frac{1}{\left( \frac{1}{H}+s\right)} - \frac{\Phi_1(\dot{s})}{\left( \frac{1}{H}+s\right)}
        - \dot{s}\Phi_2(\dot{s}) ; \label{eq:p2bmbcb}\end{array}\right.\\
\lefteqn{\mbox{at} \; t = t_0} \hspace{1.5in} &&  h_0=H_i \; \text{for $s_i\leq x\leq 1$},
\hskip0.25cm  {s} = s_i  . \label{eq:p2bicb}
\end{eqnarray}
This problem in the case slowly varying diffusivity case when $\Phi_1=0=\Phi_2$ has been considered by \cite{Hagan} in the context of binary alloy oxidation. However, the more general statement has not received attention.

\section{Discussion}\label{Discussion}

A set of reaction-diffusion equations representing the concrete carbonation process has been analysed in the limit of fast reaction - slow diffusion. Different sharp interface models in the form of generalised Stefan problems are derived, depending upon the properties of the diffusivities (of the two main species CO$_2$ and Ca(OH)$_2$) as well as the size of the relative transport parameter $\delta^2/\epsilon^2 = D_h^0 h^0/D_{c}^0 C^* \leq O(1)$. These two considerations determine the type of sharp interface Stefan and kinetic conditions. A one-phase model results for $\delta \ll \epsilon$ and a two-phase model when $\delta=O(\epsilon)$. The sharp interface kinetic conditions are determined by the reaction-diffusion equations within the thin reaction zone, thus capturing features of the model on smaller length scales (microscales).  The resulting sharp interface model considered as a macroscale model thus contains microscale information, which we argue is more appropriate than  simply stating an empirical kinetic condition.

The issue of which diffusivity situation is appropriate i.e. one or both rapidly varying, is a modelling aspect. The choice will ultimately depend on the best fit with experimental observations for the carbonation zone.

It is worth noting the following open issues:
\begin{itemize}
\item In \cite{Hornung} (sect. 1.5 and chapter 9), one derives a class of distributed-microstructure models. Roughly speaking, it is about two-scale systems of PDEs which are coupled via micro-macro boundary conditions.  \cite{Meier_thesis} and \cite{Meier1} report on such two-scale problems where, at the micro level, a sharp-interface model is responsible for the evolution of micro-free reaction boundaries.  Can one derive via homogenization techniques our two-scale model?

\item Are there any connections between our two-scale result  and the Localized Model-Upscaling (LMU) method by Pierre Degond et al. \cite{Degond}?  LMU is typically used in the context of Boltzmann-like equations. Essentially, it consists in coupling a perturbation model and its asymptotic limit model via a transition zone. In the transition zone, the solution is decomposed into a Ômicroscopic fractionÕ (described by the perturbation problem) and a Ômacroscopic oneÕ (described by the limit problem).

\item The single scale Stefan-like problems can be investigated by means of standard methods for free-boundary problems; see e.g. \cite{Primicerio}. However, the well-posedness of the two-scale FBPs summarised in Section \ref{SoltStefanProb} is an open problem. We expect that methods developed in \cite{Ai09,MB09,Mun09_RWA} and \cite{Ai10} can be adapted to cope with the two-scale structure of our problem.

\item Once the limit problems are shown to be well-posed, the next step is to prove rigorously the passing to the limit $\epsilon\to 0$ by getting an upper bound on the convergence rate (corrector estimate). This issue is open even for the derivation (as fast-reaction limit) of the standard Stefan problem.
\item For the practical corrosion problem, the geometry of the concrete structure plays an important role. We would like to extend the asymptotics to $2D$ and $3D$ cases and explore the effect of corners on the speed of the moving reaction interface.

\item Resolving numerically in an accurate manner $2D$ and $3D$ singular-perturbation scenarios require a very good understanding of the evolution of the  singular part -- the reaction layer. We hope that the asymptotic method will help us to construct a good method to capture the {\em a priori} unknown position of the layer.

\item Using eventually the same experimental data as reported in  \cite{MBK11} or \cite{PVF}, we want to validate the two-scale model. It is unclear for the moment what is the best option: Should one use the two-scale sharp interface model (which is exact), or its $\epsilon$-approximation (which is right only within a corrector range to be established)?
\end{itemize}

\section*{Acknowledgments}
AM thanks DFG SPP1122 for an initial financial support.

\bibliography{paperbib}
\bibliographystyle{plain}

\appendix

\app{Stage I carbonisation}
\label{appA}

\subsection{Slowly varying diffusivities}

Here we complete the analysis in the case $\delta=O(\epsilon)$ considered in section \ref{D0c3} for slowly varying diffusivities. The reaction zone remains at the surface until the hydroxide within it is consumed. As such $h=O(1)$ and for this inner region we consider the scalings
\be
 x= \epsilon^{\frac{2}{p+1}} \bar{x}, \hskip 1cm c=\epsilon^{\frac{2}{p+1}}\bar{c}, \hskip 1cm h= \bar{h},
\label{eq:appAscale}
\ee
giving
\begin{gather}
\epsilon^{\frac{4}{p+1}} \mu \frac{\partial \bar{c}}{\partial t}
     - \frac{\partial}{\partial \bar{x}}\left( D_c \frac{\partial \bar{c}}{\partial \bar{x}}\right)= - \bar{c}^p \bar{h}^q, \\
\epsilon^{\frac{4}{p+1}}\frac{\partial \bar{h}}{\partial t}
     -\lambda^2 \frac{\partial}{\partial \bar{x}} \left( D_h\frac{\partial \bar{h} }{\partial \bar{x}} \right) = - \epsilon^{\frac{2}{p+1}} \bar{c}^p \bar{h}^q.
\end{gather}
Posing
\be
     \bar{c} = \bar{c}_0 + \epsilon^{\frac{2}{p+1} } \bar{c}_1 + \cdots , \hskip 1cm
         \bar{h} = \bar{h}_0 + \epsilon^{\frac{2}{p+1} } \bar{h}_1 + \cdots ,
\label{eq:appAexp}
\ee
after an initial transient $t =O(\epsilon^{\frac{4}{p+1}})$, we obtain at the leading order
\begin{gather}
 \frac{\partial}{\partial \bar{x}}\left( D_c \frac{\partial \bar{c}_0}{\partial \bar{x}}\right)=  \bar{c}_0^p \bar{h}_0^q, \hskip 1cm
  \lambda^2 \frac{\partial}{\partial \bar{x}}\left( D_h \frac{\partial \bar{h}_0}{\partial \bar{x}}\right) =  0,
\end{gather}
subject to
\begin{eqnarray}
  && \text{at $\bar{x}=0$} \hskip 1cm
      D_c\frac{\partial \bar{c}_0}{\partial \bar{x}} = -H ,
 \hskip 0.25cm D_h\frac{\partial \bar{h}_0}{\partial \bar{x}} = 0 ,  \\
  && \text{as $\bar{x} \to +\infty$} \hskip 1cm \bar{c}_0 \to 0, \hskip 0.25cm
          \bar{h}_0 \to h_0 (0,t) .
\end{eqnarray}
Consequently,
\be
    \bar{h}_0 = h_0(0,t), \hskip 0.5cm \bar{c}_0 = \left\{ \begin{array}{ll}
        \frac{H}{D_c \sigma_p(t)} \left( 1 + \frac{(p-1)}{2} \sigma_p(t) x \right)^{-\frac{2}{(p-1)}}, & p>1,   \\ [2ex]
         \frac{H}{D_c \sigma_1(t)} \exp \left( - \sigma_1(t) x \right), & p=1,  \\ [2ex]
          \left(  \left(\frac{H}{D_c \sigma_1(t)} \right)^{\frac{(1-p)}{(1+p)}} - \frac{(1-p)}{2} \sigma_1(t) x \right)^{\frac{2}{(1-p)}}, & 0 \leq p<1 ,
       \end{array} \right.
\ee
where
\[
 \sigma_p(t)= \left( \frac{2 h_0(0,t)^q}{(1+p) D_c} \left(\frac{H}{D_c} \right)^{p-1} \right)^{\frac{1}{1+p}}
\]
and setting $p=1$ gives $\sigma_1(t)$. This solution recorded for $\bar{c}_0$ being relevant when $D_c$ is constant in this region.

At the next order for the hydroxide, we have
\begin{gather}
 \lambda^2 \frac{\partial}{\partial \bar{x}}\left( D_h \frac{\partial \bar{h}_1}{\partial \bar{x}}\right)=  \bar{c}_0^p \bar{h}_0^q,
\end{gather}
with
\begin{eqnarray}
  && \text{at $\bar{x}=0$} \hskip 1cm
      D_h\frac{\partial \bar{h}_1}{\partial \bar{x}} = 0 ,
\end{eqnarray}
giving
\be
     \lambda^2 D_h\frac{\partial \bar{h}_1}{\partial \bar{x}} = D_c\frac{\partial \bar{c}_0}{\partial \bar{x}} + H .
\label{eq:h1bar}
\ee

The leading order outer 2 problem is thus the fixed domain problem
\begin{eqnarray}
\lefteqn{\mbox{in} \; 0 < x < 1,  t > 0}  \hspace{1.5in} &&
   \frac{\partial  h_0}{\partial t} = \lambda^2 \frac{\partial}{\partial x} \left( D_h \frac{\partial h_0}{\partial x} \right) ;  \\
\lefteqn{\mbox{at} \; x = 0} \hspace{1.5in} &&  \lambda^2 D_h \frac{ \partial h_0 }{\partial x} = H; \label{eq:fdoutmatch} \\
\lefteqn{\mbox{at} \; x = 1} \hspace{1.5in} && D_h\frac{\partial h_0}{\partial x}
= 0  ; \\
\lefteqn{\mbox{at} \; t = 0} \hspace{1.5in} && h_0=h_i \; \text{for $0\leq x\leq 1$},
\end{eqnarray}
the condition (\ref{eq:fdoutmatch}) following from matching with the above two term inner solution (\ref{eq:appAexp}) and using (\ref{eq:h1bar}). In the case when $D_h$ and $h_i$ are constants, we have the explicit solution
\be
  h_0(x,t) = h_i  - H t + \frac{H}{\lambda^2 D_h} \left( \frac{x}{2} (2-x) + \sum_{n=0}^{\infty} f_n \exp(-\lambda^2 D_h n^2 \pi^2 t) \cos(n\pi x) \right),
\label{eq:h0fd}
\ee
where the fourier coefficients are
\be
     f_0 = \frac{1}{3}, \hskip 1cm f_n = \frac{2}{n^2\pi^2} \;\; \text{for $n\geq 1$}.
\ee
In the semi-infinite case, the corresponding solution is
\be
    h_0(x,t) = h_i - \frac{H}{\lambda D_h^{\frac{1}{2}}} \int_{0}^{t} \frac{\exp(-x^2/4\lambda^2D_h s )}{\sqrt{\pi s}} ds,
\label{eq:infsol}
\ee
We denote by $t=t_0$ the time at which the leading order hydroxide concentration falls to zero at the reaction zone i.e. $h_0(0,t)=0$, the resulting concentration profile then being $H_i(x)=h_0(x,t_0)$. In the finite domain case, (\ref{eq:h0fd}) gives a transcendental equation for $t_0$,
\be
     H t_0 = h_i + \frac{H}{\lambda^2 D_h} \sum_{n=0}^{\infty} f_n \exp(-\lambda^2 D_h n^2 \pi^2 t_0) ,
\ee
with then
\be
  H_i(x) =   h_i  - H t_0 + \frac{H}{\lambda^2 D_h} \left( \frac{1}{2} x(2-x) + \sum_{n=0}^{\infty} f_n \exp(-\lambda^2 D_h n^2 \pi^2 t_0) \cos(n\pi x) \right).
\ee
In the semi-infinite domain case, (\ref{eq:infsol}) gives explicitly
\be
     t_0 = \frac{\pi \lambda^2D_h h_i^2}{4H^2}, \hskip 1cm H_i(x) = h_i - \frac{H}{\lambda D_h^{\frac{1}{2}}} \int_{0}^{t_0} \frac{\exp(-x^2/4\lambda^2D_h s )}{\sqrt{\pi s}} ds.
\ee

Since the leading order inner solution for the hydroxide $\bar{h}_0$ does not vary spatially within the inner region, this Stage I carbonation persists until the leading order outer 2 solution falls to zero at the reaction zone. Subsequently Stage II carbonation takes place as described
in section \ref{D0c3}.

The case $H=\infty$ is different from the above, since $c$ as well as $h$ are now order 1 in the reaction zone.

\subsection{Single rapidly varying diffusivity $D_c$}

For the single rapidly varying CO$_2$ diffusivity considered in section \ref{secd=eF1}, a stage I carbonation regime takes place analogous to the slowly varying case. The C0$_2$ diffusivity is now small in the reaction zone, consequently the scalings (\ref{eq:appAscale}) change to
\be
 x= \epsilon^{\frac{4}{p+1}} \bar{x}, \hskip 1cm c=\epsilon^{\frac{4}{p+1}}\bar{c}, \hskip 1cm h= \bar{h}, \hskip 1cm D_c= \epsilon^2 \bar{D}_c,
\label{eq:D1appAscale}
\ee
and the expansion modified to
\be
     \bar{c} = \bar{c}_0 + \epsilon^{\frac{4}{p+1} } \bar{c}_1 + \cdots , \hskip 1cm
         \bar{h} = \bar{h}_0 + \epsilon^{\frac{4}{p+1} } \bar{h}_1 + \cdots .
\label{eq:D1appAexp}
\ee
The rest of the analysis remains similar, with $D_c$ replaced by $\bar{D}_c$ which we note for the specific form (\ref{eq:DcArr}) is $\bar{D}_c= \nu_1/h_0(0^+,t)^{q}$. This pertains until $h_0(0^+,t)$ falls $O(\epsilon^{2/q})$, the waiting time $t_0$ being the same in this case.

\app{The inner 1 asymptotics}
\label{appB}

We complete here the analysis of the inner 1 region in section \ref{D1c1} for the parameter regime $0 < q \leq 1$.

In the case $q=1$, this inner 1 region is the same size as the inner inner region, although located in its far-field. We scale using
\be
 x=s(t) + \epsilon^{ 2 } (S(t) + \hat{x}), \hskip 1cm c= \hat{c}, \hskip 1cm h= \epsilon^{2}\hat{h},
\label{eq:D1c1scale2b}
\ee
where $1 \ll S(t) \ll 1/\epsilon^2$. The governing equations become
\begin{gather}
 \epsilon^{ 4 } \mu \frac{\partial \hat{c}}{\partial t} -  \epsilon^{ 2 } \mu (\dot{s}+\epsilon^2 \dot{S})\frac{\partial \hat{c}}{\partial \hat{x}}
- \frac{\partial}{\partial \hat{x}}\left( D_c \frac{\partial \hat{c}}{\partial \hat{x}}\right) = - \epsilon^{4 }\hat{c}^p \hat{h}^q, \label{eq:D1c1ige1b} \\
 \epsilon^{ 2 } \frac{\partial \hat{h}}{\partial t}
  -  (\dot{s}+\epsilon^2 \dot{S})\frac{\partial \hat{h}}{\partial \hat{x}}
 - \lambda^2  \frac{\partial}{\partial \hat{x}} \left( D_h\frac{\partial \hat{h} }{\partial \hat{x}} \right) = - \hat{c}^p \hat{h}^q. \label{eq:D1c1ige2b}
\end{gather}
Posing (\ref{eq:D1c1in1exp}) with $q=1$, we obtain (\ref{eq:D1c1ileadge}) with the equation for $\hat{h}_0$ being
\be
 \dot{s}\frac{\partial \hat{h}_0}{\partial \hat{x}} + \lambda^2  \frac{\partial}{\partial \hat{x}} \left( D_h\frac{\partial \hat{h}_0 }{\partial \hat{x}} \right) =  \hat{c}_0^p \hat{h}_0^q. \label{eq:D1c1ileadgeb}
\ee
After matching to outer 1 and the inner inner we again obtain (\ref{eq:D1c1insol}) with $\hat{h}_0$  and $S(t)$ (in the case $D_h$ constant) given by
\be
  \hat{h}_0 =  \exp \left( \sigma(t) \hat{x} \right) , \hskip 1cm S(t) \sim \frac{2}{q \sigma(t)} \log(\epsilon),
\ee
where
\[
\sigma(t) = \frac{-\dot{s} + (\dot{s}^2 + 4 \lambda^2 D_h c_0(s^-,t)^p)^\frac{1}{2}}{2\lambda^2 D_h}  .
\]

In the case $0<q<1$, the inner 1 region is smaller than the inner inner. We scale using
\be
 x=s(t) + \epsilon^2 \left( S(t) + \epsilon^{ \frac{1-q}{q}} \hat{x} \right), \hskip 1cm c= \hat{c}, \hskip 1cm h= \epsilon^{\frac{2}{q}}\hat{h},
\label{eq:D1c1scale2c}
\ee
where $S(t) \leq 0$ can be taken as the location at which the inner inner solution $\bar{h}_0$  first becomes zero. We thus obtain
\begin{gather}
 \epsilon^{ \frac{2(1+q)}{q}} \mu \frac{\partial \hat{c}}{\partial t} -  \epsilon^{ \frac{(1+q)}{q}} \mu (\dot{s}+\epsilon^2S(t) )\frac{\partial \hat{c}}{\partial \hat{x}}
- \frac{\partial}{\partial \hat{x}}\left( D_c \frac{\partial \hat{c}}{\partial \hat{x}}\right) = - \epsilon^{ \frac{2(1+q)}{q} }\hat{c}^p \hat{h}^q, \label{eq:D1c1ige1c} \\
   \epsilon^{ \frac{2}{q}}  \frac{\partial \hat{h}}{\partial t}
  -   \epsilon^{\frac{1-q}{q}} (\dot{s}+\epsilon^2 S(t)) \frac{\partial \hat{h}}{\partial \hat{x}}
 - \lambda^2  \frac{\partial}{\partial \hat{x}} \left( D_h\frac{\partial \hat{h} }{\partial \hat{x}} \right) = - \hat{c}^p \hat{h}^q. \label{eq:D1c1ige2c}
\end{gather}
Posing
\be
\hat{c} \sim \hat{c}_0 + \epsilon^{\frac{(1+q)}{q} } \hat{c}_1, \hskip 1cm \hat{h} \sim \hat{h}_0,
\label{eq:D1c1in1expc}
\ee
we again obtain (\ref{eq:D1c1ileadge}) with the equation for $\hat{h}_0$ now being
\be
  \lambda^2  \frac{\partial}{\partial \hat{x}} \left( D_h\frac{\partial \hat{h}_0 }{\partial \hat{x}} \right) =  \hat{c}_0^p \hat{h}_0^q. \label{eq:D1c1ileadgec}
\ee
Matching to outer 1 and the inner inner gives (\ref{eq:D1c1insol}) with $\hat{h}_0$  (in the case $D_h$ constant) given by
\be
  \hat{h}_0 = \left\{ \begin{array}{ll}
         \left( \frac{(1-q)^2c_0(s^-,t)^p}{2(1+q)\lambda^2 D_h}  \hat{x}^2 \right)^{\frac{1}{1-q}} & \text{for $\hat{x}>0$}, \\
         0 &   \text{for $\hat{x}<0$}.
    \end{array} \right.
\ee

\end{document}